\documentclass[conference]{IEEEtran}
\IEEEoverridecommandlockouts
\usepackage{cite}

\usepackage{tikz} 
\usepackage{listings}
\usepackage{enumitem}
\usepackage{graphicx}
\usepackage{multicol, latexsym}
\usepackage{blindtext}
\usepackage{subcaption}
\usepackage{caption}
\usepackage{longtable}

\usepackage{adjustbox}  

\usepackage{csquotes}
\usepackage{amsfonts}
\usepackage{amsmath}
\usepackage{amsthm}
\usepackage{amssymb}
\usepackage{algorithm}
\usepackage{tabularx}
\usepackage{capt-of,lipsum} 

\usepackage{listings}
\usepackage[draft]{hyperref}
\usepackage{comment}
\usepackage{diagbox}
\usepackage{pdflscape}
\usepackage{booktabs}
\usepackage{makecell}
\usepackage{balance}

\usepackage{array}

\usepackage{dirtytalk}
\usepackage{lipsum}
\usepackage{url}

\usepackage{mathtools, nccmath}

\usepackage{siunitx}
\DeclareSIUnit[number-unit-product = { }] \dBm{dBm}

\usepackage{algpseudocode}
\usepackage{algorithm}

\usepackage{listing-styles}

\algnewcommand\algorithmicforeach{\textbf{for each}}
\algdef{S}[FOR]{ForEach}[1]{\algorithmicforeach\ #1\ \algorithmicdo}
\usepackage[most]{tcolorbox}
\usepackage{multirow}

\definecolor{myblue}{rgb}{0.09,0.20,0.34}
\definecolor{mygreen}{rgb}{0,0.6,0}
\definecolor{mygray}{rgb}{0.98,0.98,0.98}
\definecolor{myorange}{rgb}{0.92,0.49,0.34}
\definecolor{mywhite}{rgb}{1.0,1.0,1.0}

\definecolor{NMR}{RGB}{255,255,86}
\definecolor{MRA}{RGB}{255,231,27}
\definecolor{MRD}{RGB}{178,178,178}
\definecolor{MRIP}{RGB}{188,172,0}
\definecolor{MRS}{RGB}{161,207,106}
\definecolor{NA}{RGB}{228,60,52}
\definecolor{AI}{RGB}{255,255,86}
\definecolor{RMR}{RGB}{162,4,21}
\definecolor{RMD}{RGB}{178,178,178}
\definecolor{RMA}{RGB}{255,231,27}
\definecolor{RMIP}{RGB}{188,172,0}
\definecolor{RMS}{RGB}{161,207,106}

\newcommand{\ie}{\textit{i}.\textit{e}.,\ }
\newcommand{\eg}{\textit{e}.\textit{g}.,\ }
\newcommand{\cf}{\textit{c}\textit{f}.\ }

\newsavebox{\mybox}

\usepackage{pifont}

\usepackage{censor}
\usepackage{wasysym}

\usepackage[scaled]{helvet}
\usepackage[T1]{fontenc}
\usepackage{helvet}

\usepackage{geometry}
\begin{document}

\title{IHearYou: Linking Acoustic Features to\\DSM-5 Depressive Behavior Indicators}


\DeclareRobustCommand*{\IEEEauthorrefmark}[1]{%
  \raisebox{0pt}[0pt][0pt]{\textsuperscript{\footnotesize #1}}%
}

\author{\IEEEauthorblockN{Jonas Länzlinger\IEEEauthorrefmark{1},
Katharina O.E. Müller\IEEEauthorrefmark{2},
Burkhard Stiller\IEEEauthorrefmark{2},
Bruno Rodrigues\IEEEauthorrefmark{1}
}

\IEEEauthorblockA{\IEEEauthorrefmark{1}Embedded Sensing Group ESG, School of Computer Science SCS, 
University of St. Gallen HSG, Switzerland\\
} 
\IEEEauthorblockA{\IEEEauthorrefmark{2}Communication Systems Group CSG, Department of Informatics IfI, 
University of Zurich UZH, Switzerland\\
}
E-mail:jonas.laenzlinger@student.unisg.ch¦[stiller,mueller]@ifi.uzh.ch¦bruno.rodrigues@unisg.ch
}


\maketitle

\begin{abstract}
Depression affects over millions people worldwide, yet diagnosis still relies on subjective self-reports and interviews that may not capture authentic behavior. We present \textit{IHearYou}, an approach to automated depression detection focused on speech acoustics. Using passive sensing in household environments, \textit{IHearYou} extracts voice features and links them to DSM-5 (Diagnostic and Statistical Manual of Mental Disorders) indicators through a structured Linkage Framework instantiated for Major Depressive Disorder. The system runs locally to preserve privacy and includes a persistence schema and dashboard, presenting real-time throughput on a commodity laptop. To ensure reproducibility, we define a configuration-driven protocol with False Discovery Rate (FDR) correction and gender-stratified testing. Applied to the DAIC-WOZ dataset, this protocol reveals directionally consistent feature–indicator associations, while a TESS-based audio streaming experiment validates end-to-end feasibility. Our results show how passive voice sensing can be turned into explainable DSM-5 indicator scores, bridging the gap between black-box detection and clinically interpretable, on-device analysis.
\end{abstract}


\section{Introduction}
\label{sec:introduction}

Depression is one of the most prevalent mental health disorders, with rates continuing to rise across all age groups \cite{who2023depression, Remes2021DeterminantsDepression}. The consequences are especially severe for children and adolescents, where early symptoms are often overlooked or untreated \cite{ahrq2022child}. Often, depression is diagnosed together with one or more comorbidities such as substance use, panic disorder, and anxiety disorder \cite{Remes2021DeterminantsDepression}. In clinical practice, diagnosis is typically based on structured criteria such as the DSM-5 (Diagnostic and Statistical Manual of Mental Disorders) and self-report questionnaires like the Patient Health Questionnaire (PHQ) \cite{americanpsychiatricassociation2022dsm5, kroenke2001phq9} and, while these methods are effective in many contexts, they are susceptible to recall errors, social desirability bias, and subjectivity \cite{baumeister2007behavior}. 

As access to mental health support remains largely limited to privileged groups and the strain on health care systems continuously grows \cite{who2023depression}, the critical question has to be asked: \textit{"How do we find innovative methods to detect depression more effectively during this 'epidemic of depression'?"}. Research provides evidence (overview in Section \ref{section:relatedwork}) that recognizing problems in the early stages enables the early application of an effective and suitable treatment.

Although combining speech, text, physiology, and behavior is expected to yield the most reliable assessments \cite{zhang2024multimodal,jahan2022multimodality}, integrating heterogeneous signals on constrained, household devices is still impractical. Cloud offloading would raise major privacy risks, while local multimodal models typically exceed the resources available in household environments. Therefore, our approach concentrates on acoustic data from speech, a modality that carries clinically validated cues of depression and can be processed transparently and efficiently at the edge (\ie at a household environment).

\begin{table*}[!t]
    \centering
    \small
    \caption{Selected representative state-of-the-art.}
    \label{tab:sota-comparison}
    \resizebox{\textwidth}{!}{%
        \begin{tabular}{lccccccccc}
            \toprule
            \textbf{Work} & \textbf{Text} & \textbf{Speech} & \textbf{Vision} & \textbf{Mobile/Home} & \textbf{Net} & \textbf{Fusion} & \textbf{Setting} & \textbf{Privacy} & \textbf{Repro.}\\
            \midrule
            Sharma \emph{et al.} (2020) \cite{sharma2020twitter} & \checkmark & -- & -- & -- & -- & -- & Online & AN & PR \\
            Cacheda \emph{et al.} (2019) \cite{Cacheda2019earlyDD} & \checkmark & -- & -- & -- & -- & L & Online (early det.) & -- & PR \\
            Low \emph{et al.} (2020) \cite{Low2020keyacousticfeatures} & -- & \checkmark & -- & -- & -- & -- & Lab/clinic & -- & -- \\
            Di \emph{et al.} (2024) \cite{di2024voicepitch} & -- & \checkmark & -- & -- & -- & -- & Field/phone & -- & -- \\
            Donaghy \emph{et al.} (2024) \cite{Donaghy2024VoiceBiomarkers} & -- & \checkmark & -- & -- & -- & -- & Field & -- & -- \\
            Cohn \emph{et al.} (2009) \cite{cohn2009detecting} & -- & -- & \checkmark & -- & -- & -- & Lab & -- & -- \\
            Min \emph{et al.} (2023) \cite{Min2023detectingDepressionVideoLogs} & -- & -- & \checkmark & -- & -- & H & In-the-wild video & -- & -- \\
            Lin \emph{et al.} (2020) \cite{lin2020sensemood} & -- & -- & -- & \checkmark & -- & -- & Field (mobile) & -- & -- \\
            Shen \emph{et al.} (2017) \cite{shen2017dd} & -- & -- & -- & \checkmark & -- & -- & Field (mobile) & -- & -- \\
            Govindasamy \emph{et al.} (2021) \cite{Govindasamy2021} & -- & -- & -- & \checkmark & -- & -- & Field (wearable) & -- & -- \\
            Sadeghi \emph{et al.} (2024) \cite{sadeghi2024harnessing} & \checkmark & \checkmark & -- & -- & -- & H & Lab/online & -- & -- \\
            Atmaja \emph{et al.} (2022) \cite{trisatmaja2022bimodalvoice} & -- & \checkmark & \checkmark & -- & -- & E & Lab & -- & -- \\
            DAIC-WOZ \cite{usc_daic_woz} (2024) & -- & \checkmark & \checkmark & -- & -- & -- & Lab & -- & DA \\
            \midrule
            \textbf{IHearYou} & -- & \checkmark & -- & \checkmark & -- & H & Home (edge) & ED & DA+PR \\
            \bottomrule
        \end{tabular}
    }   
    \begin{flushleft}
        \footnotesize `Text''=social/self-report text; Speech''=paralinguistics; Vision''=facial/gaze/body cues; Mobile/Home''=passive signals from phone, wearables, and ambient/home devices; Net''=network-derived digital traces. Fusion: E=early, L=late, H=hybrid. Privacy: ED=edge/on-device, AN=anonymization, FL=federated. Repro.: DA=data access, CO=code, PR=protocols.
    \end{flushleft}
\end{table*}

For speech in particular, prior work shows that acoustic cues (\eg reduced pitch variability, slower speech, and longer pauses) are related to depressive states, yet most studies report aggregate accuracy without exposing which feature supports which DSM‑5 indicator or how context modifies that link \cite{Yadav2022ADD,Low2020keyacousticfeatures}. Without such a precise mapping between audio features and symptoms, outputs are hard to interpret and use in clinical workflows.

We investigate this challenge assessing whether low‑level acoustic features extracted from raw speech can be mapped to specific DSM‑5 depressive‑behavior indicators in a transparent and resource‑aware way. Concretely, we test the following hypotheses: (H1) reduced pitch variability and (H2) longer pauses are each positively associated with the DSM‑5 indicators \emph{psychomotor retardation} and \emph{diminished ability to think or concentrate}, after adjusting for channel, language, age, and sex; (H3) lower energy dynamics and (H4) slower speech rate show the same direction of association. The null hypothesis for each is that no association remains after adjustment and correction for multiple testing.

However, building this mapping is not a trivial task. Acoustic cues (\eg reduced pitch variability and longer pauses within a speech) can be present in multiple processes at once and are sensitive to various factors, such as channel conditions, language, age, and sex. For instance, such cues are consistent with the DSM-5 indicators psychomotor retardation and diminished ability to think/concentrate \cite{AdamsQuackenbush2019,Low2020keyacousticfeatures,di2024voicepitch,Donaghy2024VoiceBiomarkers,Bains2023}. Simplifying these many-to-many relations requires a structured mapping that controls for confounders such as age, sex, and channel conditions, remains efficient enough for edge execution, and produces interpretable rationales at the case level, making its conclusions explainable to DSM-5 indicators.

\emph{IHearYou} links measurable acoustic features to DSM‑5 \cite{americanpsychiatricassociation2022dsm5} depression indicators through a transparent, verifiable specification that runs locally at a home network, preserving privacy while enabling near‑real‑time analysis under tight memory, latency, and energy budgets \cite{linthicum2023ubiquitous,dilmegani2025tinyml}. The design is modular: when available, additional behavioral signals from mobile, wearable, or ambient sensors can be fused to strengthen the evidence without changing the on‑device privacy model. The main contributions of this paper are summarized as follows:

\begin{itemize}
  \item \textbf{Indicator-level mapping}: we introduce a DSM-5–aligned Linkage Framework that connects concrete acoustic features to depressive-behavior indicators through explicit, testable mappings and scoring rules.  
  \item \textbf{Edge design}: we implement a local, privacy-preserving pipeline with a persistence schema and dashboard, and demonstrate sustained real-time throughput on a commodity laptop under continuous audio load.  
  \item \textbf{Reproducible empirical evidence}: we provide a configuration-driven protocol with False Discovery Rate (FDR) correction and gender-stratified analysis, applied to DAIC-WOZ \cite{usc_daic_woz} showing directionally consistent feature–indicator associations, and to TESS \cite{TESS_dataset} to validate streaming feasibility.  
\end{itemize}

\section{Related Work}
\label{section:relatedwork}

Depression detection has been explored across text, speech, vision, and mobile sensing, increasingly with multimodal learning. Below we review these lines of work outlining key selected work (\cf Table \ref{tab:sota-comparison}) and identify open gaps.

\subsection{Text from Social Media and Self-Reports}
\label{section:relatedwork:text}
Text-based methods analyze posts, captions, and self-reports to infer depressive signals at the lexical, syntactic, and semantic levels. Surveys summarize progress and datasets, emphasizing early-detection protocols and evolving deep models \cite{mao2023systematic,Yadav2022ADD}. Representative systems focus on platform-specific corpora (\eg Twitter) and demonstrate improvements via transformer-based embeddings and temporal patterns \cite{sharma2020twitter,Cacheda2019earlyDD}. Studies often rely on automatic preprocessing (tokenization, normalization, spelling correction) \cite{Chiong2021smn,norvig_spell_corrector,bird2009nltk}. However, labels frequently come from weak proxies (self-disclosure, subreddit membership), and the domain/temporal shift across platforms limits deployment beyond the Web.

\begin{figure*}[h]
    \centering
    \includegraphics[width=\textwidth]{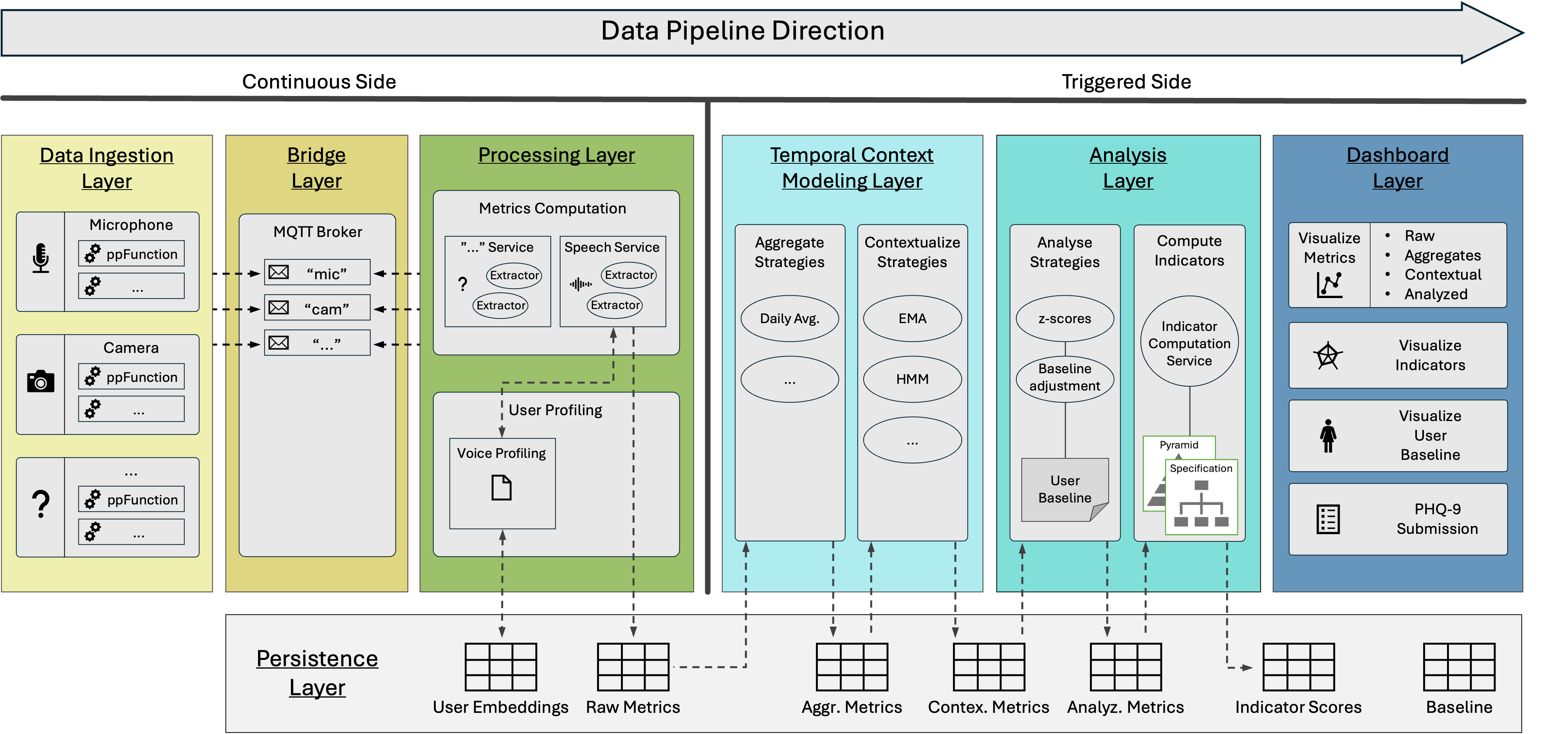}
    \caption{High‑level system architecture.}
    \label{fig:high-level-architecture}
\end{figure*}

\subsection{Speech and Paralinguistics}
\label{section:relatedwork:speech}
Speech carries prosodic and spectral markers associated with depressive symptoms. Reviews cover acoustic descriptors and paralinguistic pipelines, along with limitations of small, lab-collected datasets \cite{cummins2015review,Low2020keyacousticfeatures}. Recent work explores ``voice biomarkers’’ including pitch dynamics and phone-based short recordings \cite{di2024voicepitch,Donaghy2024VoiceBiomarkers}. Already in 1921, Kraepelin described depressive voices as those of patients speaking in a low and slow voice. He notes that such speech is often characterized by hesitation, monotony, whispering, and occasional stuttering. Furthermore, patients may abruptly stop mid-sentence \cite{kraepelin1921manic}. 

An early work by Cohn \textit{et al.} compared depression with affective computing and found that particular vocal biomarkers, such as voice prosody, can offer valuable insights into the individual's mental health state. They leveraged pitch extraction methods to measure vocal expression and logistic regression classifiers, resulting in an overall 79\% classification accuracy \cite{cohn2009detecting}. 

The work by Huang \textit{et al.} represents another great milestone. They used \textit{wav2vec 2.0} as a voice-based pre-training model for the feature extraction \cite{huang2024depression}. The speech embeddings then were used in a classification model and achieved on the popular DAIC-WOZ dataset \cite{usc_daic_woz} an outstanding classification accuracy of 96.49\% \cite{huang2024depression}.

\subsection{Vision and Behavioral Cues}
\label{section:relatedwork:vision}
Visual signals such as facial expressivity, gaze, and head motion, have been leveraged to detect depressive states. For example, controlled interviews to in-the-wild video logs \cite{cohn2009detecting,Min2023detectingDepressionVideoLogs}. Methods used span handcrafted features to deep spatiotemporal encoders, yet privacy risks are salient and most data are collected under constrained protocols, which limits longitudinal, real-home use.

\subsection{Mobile, Wearable, and Home/Network Sensing}
\label{section:relatedwork:mobile}
Digital phenotyping uses passively collected signals (\eg as activity, sleep, mobility, app/network use) to approximate behaviors associated with mood \cite{lin2020sensemood,shen2017dd}. Previous studies report correlations between passives and validated scales, as well as model-based screening pipelines \cite{Govindasamy2021,Kong2022,VANDANA2023100587,LIU202544,tufail2023dd}. These systems typically emphasize phones and wearables; few work integrate ambient/home sensors or network-layer traces in a unified, privacy-preserving architecture.

\subsection{Multimodal Fusion}
\label{section:relatedwork:multimodality}
Multimodal learning aims to combine complementary cues for robustness and stability. Prior work demonstrates gains by fusing speech–text or audio–visual signals for affect and depression detection \cite{sadeghi2024harnessing,trisatmaja2022bimodalvoice,zhang2024multimodal}. Fusion strategies include early (feature-level), late (decision-level), and hybrid/attention-based designs. While these studies show consistent improvements, many rely on lab datasets and cloud inference, with limited attention to on-device processing, privacy, and reproducibility (shared protocols, ablations, and release artifacts).



\section{System Design}
\label{sec:design}

In recent years, the most dominant approach for detecting depressive behaviors has relied on deep learning, ML techniques employing NNs \cite{cummins2015review}. There exists a substantial trade-off between model performance and their explainability: ML models are capable of revealing undiscovered non-linear relationships and latent structures in data that are not necessarily comprehensible by humans \cite{herm2023mlexplainability}. As Fried \textit{et al.} suggests, moving away from total depression scores to understanding symptoms at the symptom level is essential for appropriate treatment \cite{fried2025depressionsumscores}.

The high-level architecture (\cf Figure~\ref{fig:high-level-architecture}) describes key components and the respective, service granularity, and service communication. Data flow follows a left-to-right structure, reflecting the progression through the data pipeline organized into a \textit{continuous side} and a \textit{triggered side}:
\begin{itemize}
    \item \textbf{Continuous side}: processes the streamed data as it arrives in a near-real-time manner.
    \item \textbf{Triggered side}: only progresses through the individual pipeline steps when the user triggers the computation manually.
\end{itemize}

\begin{figure}[t]
    \centering
    \includegraphics[width=1.0\linewidth]{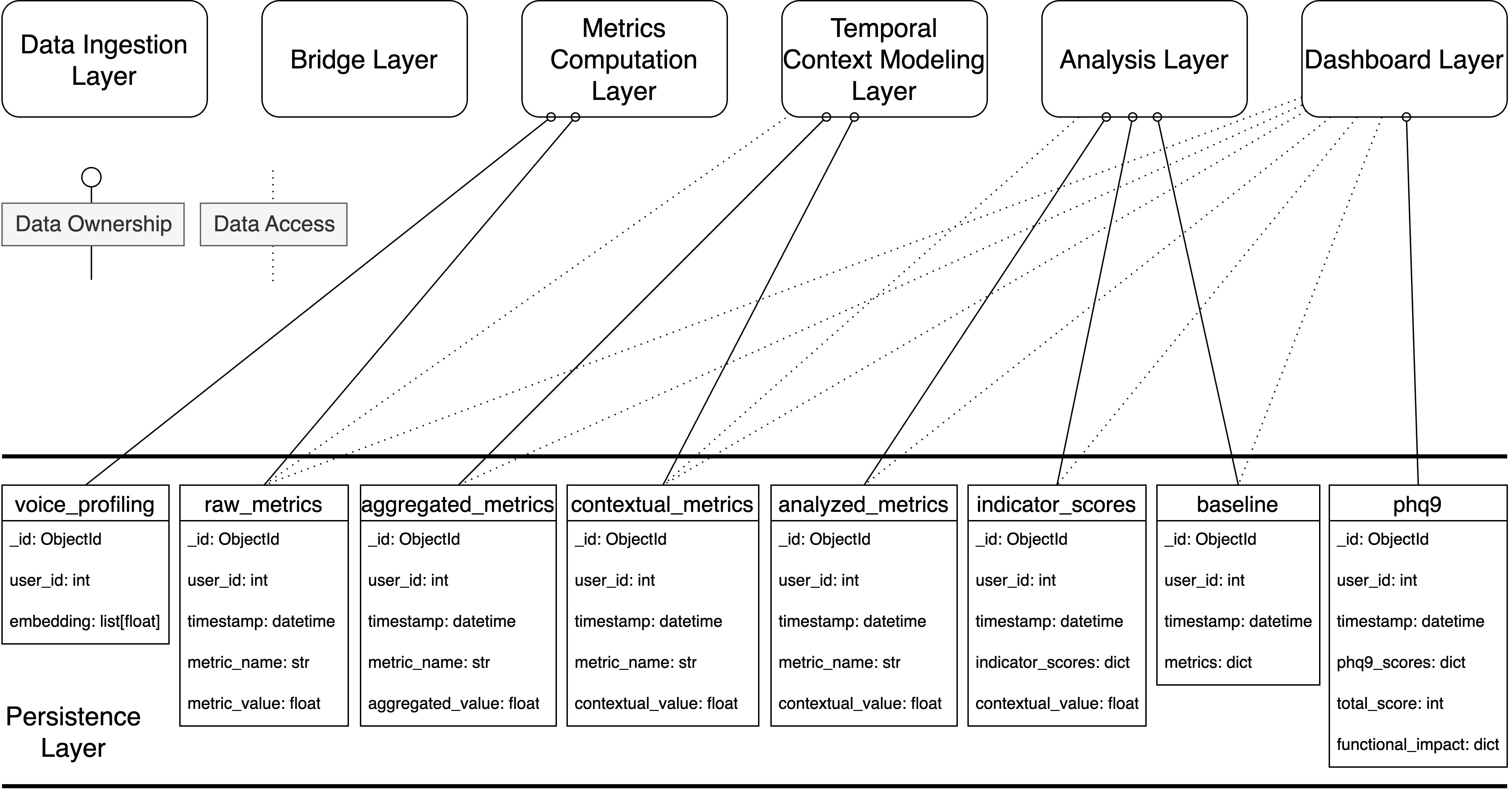}
    \caption{Persistence layer overview: database collections model}
    \label{fig:database_model}
\end{figure}

The \textbf{data ingestion layer} standardizes how heterogeneous sensor data enter the system through a collect–filter–transport process. Built on abstract device classes, it supports scalable integration of modalities while enabling edge filtering. In this work, audio is the primary input, where microphones capture speech and SileroVAD \cite{SileroVAD} removes silence and noise before forwarding voiced segments to the bridge layer.

The \textbf{bridge layer} normalizes audio levels, unifying formats, and harmonizing metadata for consistent measurements across devices. The \textbf{processing layer} extracts low-level descriptors (\eg LLD; $F_0$, intensity, jitter, shimmer, spectral bands) on 25\,ms frames with 10\,ms hops, applying VAD to remove non-speech, and discarding low-quality frames. These are aggregated into high-level descriptors (HLD) over fixed 10\,s windows using functionals (mean, variance, range, spectral modulation energy) for robustness.

HLDs and quality metadata are stored in the \textbf{persistence layer} for reproducibility. At this layer, data ownership and data access are designed towards one single database (\cf Figure \ref{fig:database_model}) instance containing multiple collections, where, respectively, one service is responsible for the writing operations, but multiple services can access the data simultaneously. 

In the collections that store metric data that has been captured and processed by the system (\textit{raw\_metrics}, \textit{aggregated\_metrics}, \textit{contextual\_metrics}, and \textit{analyzed\_metrics}), are stored as one record per metric. This enables flexible querying and efficient writing operations due to the granular nature of the records. 

The \textbf{temporal context layer} aligns features across windows and applies indicator-specific smoothing: gradual biomarkers (\eg depressed mood) integrate over time, abrupt ones (\eg psychomotor agitation) respond rapidly. Lastly, the \textbf{analysis layer} maps features to biomarkers and indicators, applies weights and directionality, and generates continuous indicator scores. Scores are binarized by thresholds, persistence, and presence criteria, then combined under the DSM-5 rule to produce the diagnosis. Above L1, all processing runs on-device, with only aggregated scores and non-identifiable metadata transmitted to uphold privacy-by-design.

\subsection{Linkage Framework (LF)}

The LF establishes how low-level acoustic measurements are transformed into interpretable evidence of depression. It formalizes a layered mapping process (to ease the comprehension of what is realized at each stage), that bridges raw audio data with the symptom indicators defined in DSM-5:
\begin{itemize}
    \item \textbf{Measurement}: frame-level $F_0$, intensity, jitter, shimmer. 
    \item \textbf{Feature}: functionals such as $F_0$ range, intensity variance, spectral slope, harmonics-to-noise ratio (HNR), cepstral peak prominence (CPP). 
    \item \textbf{Biomarker}: monopitch, monoloudness, breathiness, slowed speech). 
    \item \textbf{Indicator}: DSM-5 symptom items.
    \item \textbf{Analysis}: counting rule over indicators.
\end{itemize}

\begin{figure}[H]
    \centering
    \includegraphics[width=1.0\linewidth]{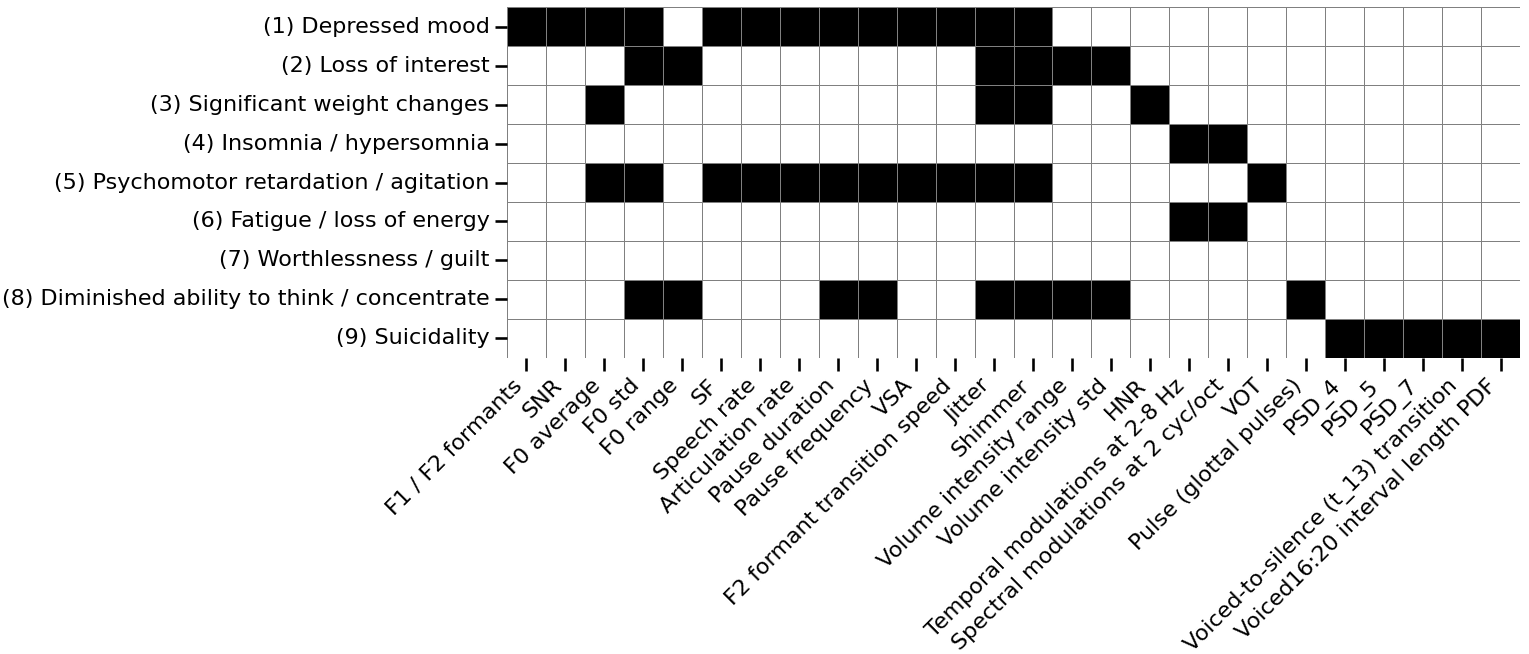}
    \caption{Linking feature intersections – matrix plot.}
    \label{fig:feature_intersections_2}
\end{figure}

Each layer has defined inputs and outputs, ensuring modularity: features can be extended without altering the clinical logic, and decision rules can evolve without re-engineering the feature pipeline. Importantly, mappings are \textbf{many-to-many}: one feature may inform multiple indicators, while each indicator is supported by multiple features. This redundancy is deliberate, as overlapping evidence increases robustness against noise and missing data. As Figure \ref{fig:feature_intersections_2} illustrate, there are several intersections between voice features and indicators, which means that individual features can be used multiple times while they are still computed only once, thereby enabling transfer of information across features and supporting a more efficient system. 

For example, reduced $F_0$ range supports both depressed mood and loss of interest, while abnormal articulation rate contributes to psychomotor retardation and diminished concentration. By structuring these links in a machine-readable format, the framework preserves transparency, reproducibility, and clinical interpretability.

\begin{table}
\centering
\caption{DSM-5 symptom indicators for MDD \cite{americanpsychiatricassociation2022dsm5}.}
\label{tab:dsm5-symptoms}
\begin{tabular}{cl}
\toprule
\textbf{ID} & \textbf{Symptom Indicator} \\
\midrule
(1) & Depressed mood \\
(2) & Loss of interest or pleasure (anhedonia) \\
(3) & Significant weight change or appetite disturbance \\
(4) & Insomnia or hypersomnia \\
(5) & Psychomotor retardation or agitation \\
(6) & Fatigue or loss of energy \\
(7) & Feelings of worthlessness or excessive guilt \\
(8) & Diminished ability to think, concentrate, or indecisiveness \\
(9) & Recurrent thoughts of death or suicidality \\
\bottomrule
\end{tabular}
\end{table}

\subsection{Feature Mapping Specification}

The next part of the \textit{LF} is the specification of the selected features, \ie their associative relationship and their individual contribution to the calculation of the DSM-5 indicator scores. This specification defines how each lower-level metric is interpreted and the intermediate abstraction it represents (biomarker). Not all acoustic features contribute equally or in the same way to the calculation of the DSM-5 indicator scores. 

The associative relationship between the voice features and the DSM-5 indicators can be either \textit{gradual} or \textit{non-linear}. A gradual relationship with a given direction \textit{positive}, \textit{negative}, or \textit{both}, simply indicates that a shift of the voice feature influences the associated DSM-5 indicator in terms of the defined direction. Otherwise, an associative relationship might be \textit{non-linear}. This means that the correlation is exponential, logarithmic, stair-shaped with some kind of triggering threshold effects, or even more complex. Table \ref{tab:unified_feature_mapping} details how we classify voice features and link to DSM-5 symptom indicators (\cf Table \ref{tab:dsm5-symptoms}).

In addition, two main categories are prominent in Table \ref{tab:unified_feature_mapping}: formant features, which describe the resonant frequencies of the vocal tract (\eg F1/F2, vowel space area, transition speeds) and capture articulation precision or slurred speech; and prosodic features, which describe speech melody and rhythm (\eg pitch variability, intensity, speech rate, pauses) and reflect monotony, slowed tempo, or prolonged silences.

\begin{table}[t]
  \centering
  \caption{Acoustic feature mapping specification with descriptor/type.}
  \label{tab:unified_feature_mapping}
  \scriptsize
  \setlength{\tabcolsep}{3pt}
  \renewcommand{\arraystretch}{1.05}
  \resizebox{\columnwidth}{!}{%
  \begin{tabular}{l l l l l c l l}
    \toprule
    \textbf{Acoustic feature} & \textbf{Biomarker} & \textbf{Ind.} & \textbf{Rel.} & \textbf{Dir.} & \textbf{W} & \textbf{Desc.} & \textbf{Signal group} \\
    \midrule
    F1/F2 formants & Whispering & (1) & Gradual & +  & 1.0 & LLD & Formant \\
    SNR & Whispering & (1) & Gradual & -- & 1.0 & LLD & Source \\
    $F0_{avg}$ & Monotonous speech & (1) & Gradual & -- & 1.0 & HLD & Prosodic \\
    $F0_{std}$ & Monotonous speech & (1) & Gradual & -- & 1.0 & HLD & Prosodic \\
    SF & Monotonous speech & (1) & Gradual & -- & 1.0 & LLD & Spectral \\
    Speech rate & Slow speech & (1) & Gradual & -- & 1.0 & HLD & Prosodic \\
    Articulation rate & Slow speech & (1) & Gradual & -- & 1.0 & HLD & Prosodic \\
    Pause duration & Prolonged pauses & (1) & Gradual & +  & 1.0 & HLD & Prosodic \\
    Pause frequency & Prolonged pauses & (1) & Gradual & +  & 1.0 & HLD & Prosodic \\
    VSA & Slurred speech & (1) & Gradual & -- & 1.0 & LLD & Formant \\
    $F2$ transition speed & Slurred speech & (1) & Gradual & -- & 1.0 & LLD & Formant \\
    Jitter & Speech quality & (1) & Gradual & +  & 1.0 & LLD & Source \\
    Shimmer & Speech quality & (1) & Gradual & +  & 1.0 & LLD & Source \\
    \midrule
    Vol. intensity range & Monoloudness & (2) & Gradual & -- & 1.0 & HLD & Prosodic \\
    Vol. intensity std. & Monoloudness & (2) & Gradual & -- & 1.0 & HLD & Prosodic \\
    $F0_{range}$ & Monopitch & (2) & Gradual & -- & 1.0 & HLD & Prosodic \\
    $F0_{std}$ & Monopitch & (2) & Gradual & -- & 1.0 & HLD & Prosodic \\
    Jitter & Monoquality & (2) & Gradual & +  & 1.0 & LLD & Source \\
    Shimmer & Monoquality & (2) & Gradual & +  & 1.0 & LLD & Source \\
    \midrule
    $F0_{avg}$ & Pitch & (3) & Gradual & -- & 1.0 & HLD & Prosodic \\
    Jitter & Voice breathiness/noise & (3) & Gradual & +  & 1.0 & LLD & Source \\
    Shimmer & Voice breathiness/noise & (3) & Gradual & +  & 1.0 & LLD & Source \\
    HNR & Voice breathiness/noise & (3) & Gradual & -- & 1.0 & LLD & Source \\
    \midrule
    Temporal mod. (2--8 Hz) & Speech prosody & (4) & Gradual & both & 1.0 & HLD & Prosodic \\
    Spectral mod. (2 cyc/oct) & Speech quality & (4) & Gradual & both & 1.0 & HLD & Spectral \\
    \midrule
    Speech rate & Speech prod. speed & (5) & Gradual & -- & 1.0 & HLD & Prosodic \\
    Articulation rate & Speech prod. speed & (5) & Gradual & -- & 1.0 & HLD & Prosodic \\
    Pause duration & Speech interruption & (5) & Gradual & +  & 1.0 & HLD & Prosodic \\
    Pause frequency & Speech interruption & (5) & Gradual & +  & 1.0 & HLD & Prosodic \\
    VOT & Speech initiation & (5) & Gradual & +  & 1.0 & LLD & Prosodic \\
    $F2$ transition speed & Vocal tract constriction & (5) & Gradual & -- & 1.0 & LLD & Formant \\
    $F0_{avg}$ & Liveliness & (5) & Gradual & -- & 1.0 & HLD & Prosodic \\
    $F0_{std}$ & Liveliness & (5) & Gradual & -- & 1.0 & HLD & Prosodic \\
    SF & Liveliness & (5) & Gradual & +  & 1.0 & LLD & Spectral \\
    VSA & Articulation precision & (5) & Gradual & -- & 1.0 & LLD & Formant \\
    $F2$ transition speed & Articulation precision & (5) & Gradual & -- & 1.0 & LLD & Formant \\
    Jitter & Speech quality & (5) & Gradual & +  & 1.0 & LLD & Source \\
    Shimmer & Speech quality & (5) & Gradual & +  & 1.0 & LLD & Source \\
    \midrule
    Temporal mod. (2--8 Hz) & Speech prosody & (6) & Gradual & both & 1.0 & HLD & Prosodic \\
    Spectral mod. (2 cyc/oct) & Speech quality & (6) & Gradual & both & 1.0 & HLD & Spectral \\
    \midrule
    Jitter & Pitch instability & (8) & Gradual & +  & 1.0 & LLD & Source \\
    Shimmer & Pitch instability & (8) & Gradual & +  & 1.0 & LLD & Source \\
    Pulse (glottal pulses) & Vocal motor control & (8) & Gradual & -- & 1.0 & LLD & Source \\
    Pause duration & Voice breaks & (8) & Gradual & +  & 1.0 & HLD & Prosodic \\
    Pause frequency & Voice breaks & (8) & Gradual & +  & 1.0 & HLD & Prosodic \\
    $F0_{range}$ & Monopitch & (8) & Gradual & -- & 1.0 & HLD & Prosodic \\
    $F0_{std}$ & Monopitch & (8) & Gradual & -- & 1.0 & HLD & Prosodic \\
    Vol. intensity range & Monoloudness & (8) & Gradual & -- & 1.0 & HLD & Prosodic \\
    Vol. intensity std. & Monoloudness & (8) & Gradual & -- & 1.0 & HLD & Prosodic \\
    \midrule
    PSD\_4 & Voice production & (9) & Gradual & both & 1.0 & LLD & Prosodic \\
    PSD\_5 & Voice production & (9) & Gradual & both & 1.0 & LLD & Prosodic \\
    PSD\_7 & Voice production & (9) & Gradual & both & 1.0 & LLD & Prosodic \\
    $t_{13}$ transition & Vocalization & (9) & Gradual & +  & 1.0 & HLD & Prosodic \\
    \textit{voiced16:20} PDF & Vocalization & (9) & Gradual & -- & 1.0 & HLD & Prosodic \\
    \bottomrule
  \end{tabular}}
  \vspace{2pt}
  
  \raggedright\footnotesize
  \textit{Notes:} Dir. $\in \{+,\;--,\;\text{both}\}$ denotes directionality (positive, negative, both). Descriptors: HLD/LLD. "W." denotes weight.\\[-2pt]
\end{table}

\subsection{Computing Feature-Indicator Scores}

The last and major step in the \textit{LF} consists of the transformation of the \textit{contextual metrics} into the individual DSM-5 indicator scores. Selected features are grouped by their associative relationship to depression indicators and their respective analysis methodologies. The transformation process is divided into two parts: (1) the \textit{contextual metrics} are analyzed by quantifying the deviation of the observed value from the user's baseline, and (2) the results are evaluated based on the defined associative mappings to build the \textit{indicator scores}. 

The steps below implements the flow described in the \emph{Temporal Context Modeling Layer} and the \emph{Analysis Layer} using the mappings in Table~\ref{tab:unified_feature_mapping}. The formulation is linear in structure, and reproducible from a configuration file (\texttt{config.json}).

\noindent\textbf{Step 1: Standardization (feature level)}:
in a similar way as \cite{zhang2024multimodal}, we use z-score to normalize the features to ensure equal contribution during the computation of the individual DSM-5 indicator scores. We reuse the z‑score normalization and clip extreme deviations for robustness:
\begin{equation}
\label{eq:z1}
z_m(t)=\frac{x_m(t)-\mu_m}{\max(\sigma_m,\varepsilon)},\qquad
\end{equation}
\begin{equation}
\label{eq:z2}
z_m(t)=\mathrm{sign}\!\bigl(z_m(t)\bigr)\cdot\min\!\bigl(|z_m(t)|,\tau_m\bigr),
\end{equation}
with $\varepsilon=10^{-6}$ and a per‑metric cap $\tau_m$ (default $\tau_m=3$).

Equation \ref{eq:z1} converts raw audio metrics, such as pitch, jitter, or articulation rate, into standardized z-scores relative to the user’s baseline, making them comparable across speakers and sessions. Equation \ref{eq:z2} limits outliers by clipping the absolute value to a maximum threshold $\tau_m$, while preserving its direction. This prevents extreme deviations, \eg sudden microphone noise, unusually long pauses, or atypical pitch breaks, from overwhelming the indicator scoring.

\noindent\textbf{Step 2: Directional transform (mapping logic)}:
before building the individual indicator sums, the direction modifier is applied to the z-score, as in Equation \ref{eq:dir}:
\begin{equation}
\label{eq:dir}
\psi_{i,m}(t)=
\begin{cases}
\tilde z_m(t), & \text{if }\texttt{direction}_{i,m}=\text{positive},\\
-\tilde z_m(t), & \text{if }\texttt{direction}_{i,m}=\text{negative},\\
|\tilde z_m(t)|, & \text{if }\texttt{direction}_{i,m}=\text{both}.
\end{cases}
\end{equation}

The mapping is not arbitrary, features can correlate with a DSM-5 symptom in different ways (as specified in the \textit{Feature Mapping Specification}), \eg a higher value may indicate higher severity (positive); a lower value may indicate higher severity (negative relation); or deviations in both directions may matter (too fast or too slow speech).

\noindent\textbf{Step 3: Temporal stability (indicator level)}:
according to the DSM-5 specification, five (or more) clinical indicators must be present during the same two-week period \cite{americanpsychiatricassociation2022dsm5}. To reflect this temporal requirement in the \textit{contextual\_metrics}, we optionally smooth $S_i(t)$ with an exponential moving average (EMA), simulating the intended temporal span of the DSM-5 specification:
\begin{equation}
\label{eq:ema}
\bar S_i(t)=(1-\beta_i)S_i(t)+\beta_i\,\bar S_i(t-1),\quad \beta_i\in[0,1).
\end{equation}

Equation~\eqref{eq:ema} introduces temporal stability at the indicator level. Since DSM-5 requires symptoms to be present over a sustained period (typically two weeks), raw indicator scores $S_i(t)$ cannot be judged in isolation. The EMA smooths each trajectory by combining the current score with past values, controlled by the smoothing factor $\beta_i$. A high $\beta_i$ gives stronger weight to historical evidence (appropriate for gradual symptoms such as depressed mood), while a low $\beta_i$ makes the indicator more responsive to rapid changes (\eg psychomotor agitation). For instance, if pause duration spikes abnormally for one day, the EMA quickly damps it, while several consecutive days of slowed speech gradually push the smoothed score above threshold, mimicking DSM-5’s two-week persistence requirement.

\noindent\textbf{Step 4: DSM‑5 count‑of‑nine support signal.} To combine continuous scores with the DSM-5 decision rule, each EMA smoothed indicator trajectory $\bar S_i(t)$ is converted into a binary presence variable $B_i(t)$ by applying a threshold $\vartheta_i$:  
\begin{equation}
\label{eq:binarization}
B_i(t) = \mathbb{1}\{\bar S_i(t) \geq \vartheta_i\},
\end{equation}
where $\vartheta_i$ denotes the severity threshold for indicator $i$. Each binarized indicator $B_i(t)$ signals the presence of a symptom relative to baseline. The support signal is set when at least five of nine DSM-5 indicators are active, including either depressed mood (1) or loss of interest (2) (Eq.~\ref{eq:mdd}). Baseline values $(\mu_m,\sigma_m)$ continue to adapt over time (via PHQ-9 feedback), ensuring that symptom presence remains calibrated to the user’s evolving behavior.

Let $C(t)=\sum_{i=1}^9 B_i(t)$ and $A(t)=\mathbb{1}\{B_1(t)=1 \lor B_2(t)=1\}$. The system reports:
\begin{equation}
\label{eq:mdd}
\textsc{MDD\_support}(t)=\mathbb{1}\{C(t)\ge 5 \land A(t)=1\},
\end{equation}
matching the DSM‑5 structure \cite{americanpsychiatricassociation2022dsm5}. Clinical evaluation and exclusions are outside system scope.

Equation~\eqref{eq:mdd} operationalizes the DSM-5 rule: a diagnosis is supported when at least five symptom indicators are present simultaneously, with either depressed mood or loss of interest among them. For example, if seven indicators are above threshold but neither (1) nor (2) is active, the support signal remains $0$, while if exactly five indicators are present and one of them is (1) or (2), the signal is set to $1$.

The formulation links Layer~1–2 measurements to Layer~4 indicators through explicit, testable mappings (Figure~\ref{fig:feature_intersections_2}, Table~\ref{tab:unified_feature_mapping}). It aligns the temporal windowing of the contextual metrics with DSM‑5 persistence and exposes coverage $\gamma_i(t)$ for transparency. Baseline finetuning adapts $(\mu_m,\sigma_m)$ using PHQ‑9 feedback without changing the mapping logic, keeping the system interpretable and reproducible.

\section{Evaluation}
\label{sec:evaluation}

This section evaluates whether the Linkage Framework (LF) maps low-level acoustic features to DSM-5 depressive-behavior indicators in a way that is statistically sound, reproducible, and practical for edge execution. We first describe datasets and ground truth, then the testing protocol and metrics, present results on feature--indicator associations, and finally assess efficiency on edge hardware. We close with a discussion on implications and limitations.

\subsection{Datasets and Ground Truth}
\label{sec:evaluation:data}

We use the DAIC-WOZ corpus to test feature-indicator hypotheses because it contains speech recordings and PHQ-8 \cite{kroenke2009phq8} (proxy to PHQ-9) item responses that align with DSM-5 indicators (Table~\ref{tab:phq_mapping}). From the DAIC-WOZ dataset, a random selection of 64 participants ($n=64$) was chosen while maintaining a balanced gender distribution (50\% male, 50\% female).  

\begin{figure}[H]
    \centering
    \includegraphics[width=1.0\linewidth]{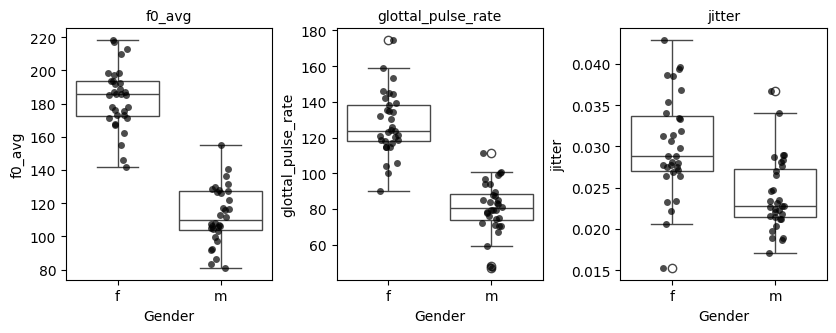}
    \caption{Distribution of selected male and female acoustic voice features.}
    \label{fig:feature_plot_mf_differences}
\end{figure}

\begin{figure}[H]
    \centering
    \includegraphics[width=1.0\linewidth]{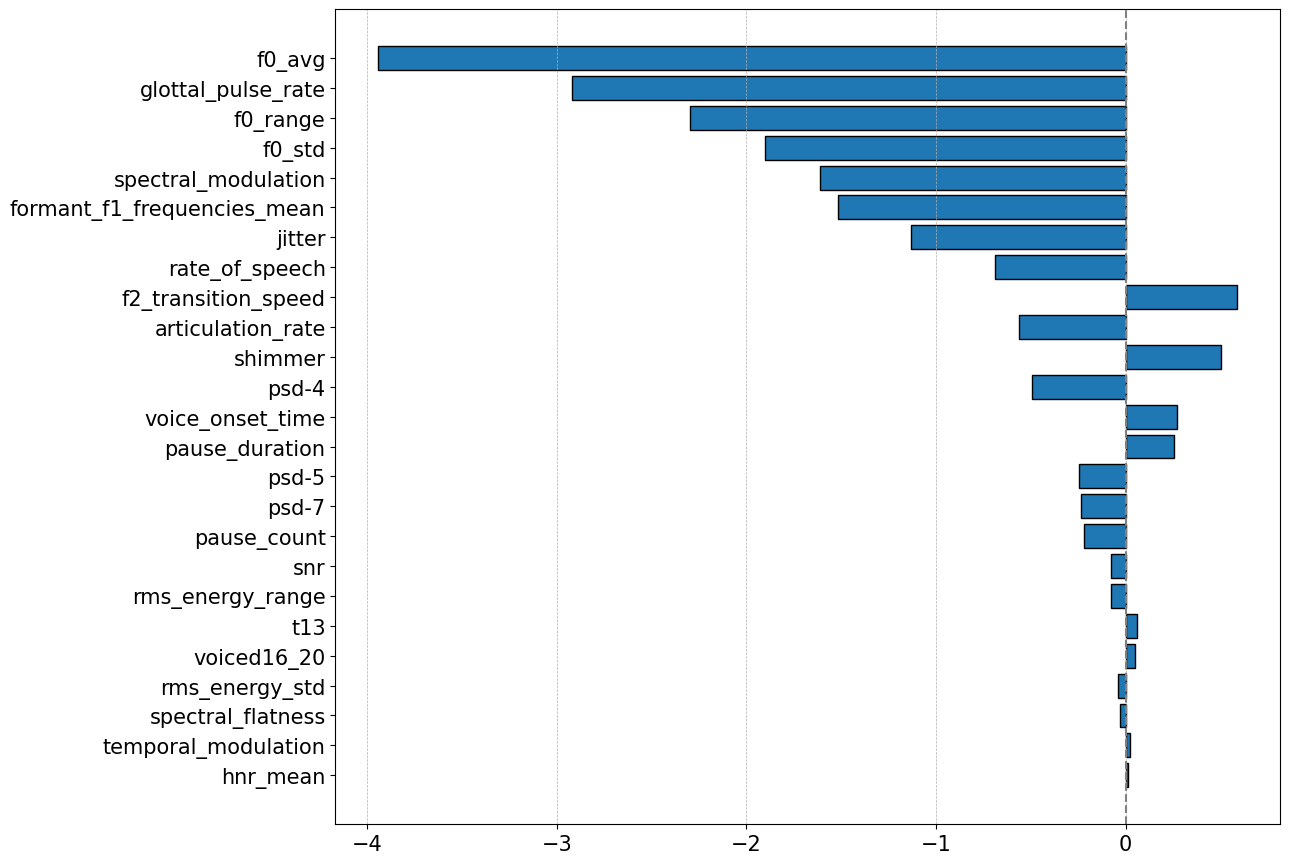}
    \caption{\textit{Cohen's d} (positive = higher in males, negative = higher in females).}
    \label{fig:cohensd_value_features}
\end{figure}

By plotting the participant data by the collected voice features, see Figure \ref{fig:feature_plot_mf_differences}, it is revealed that some of the features have differences between male and female participants, such as \textit{f0\_avg}, \textit{glottal\_pulse\_rate}, or \textit{jitter}. Figure \ref{fig:cohensd_value_features} shows the differences of each voice feature, quantified by the \textit{Cohen's d} value. Typically, if the \textit{Cohen's d} value is below $\leq\pm0.2$, the difference between two groups is considered negligible \cite{NU_CohensD_2025}. To avoid testing hypotheses where strong gender-specific effects are present, the dataset is divided into male and female groups where at least a small effect is observed.

Speech was processed with VAD, 25 ms frames/10 ms hop, and aggregated into 10 s HLDs with quality metadata for reproducibility. We then use the LF to standardize features, clip outliers, and apply directionality before computing indicator scores. 

\begin{table}[t]
\centering
\caption{PHQ item to DSM-5 indicator linkage used as ground truth in DAIC-WOZ.}
\label{tab:phq_mapping}
\begin{tabular}{p{3.3cm}p{4.3cm}}
\toprule
\textbf{PHQ-9 item} & \textbf{DSM-5 indicator}\\
\midrule
Q1: Little interest/pleasure & (2) Loss of interest \\
Q2: Feeling down/depressed & (1) Depressed mood \\
Q3: Trouble sleeping & (4) Insomnia/hypersomnia \\
Q4: Tired/low energy & (6) Fatigue \\
Q5: Poor appetite/overeating & (3) Weight/appetite \\
Q6: Feeling bad about self & (7) Worthlessness/guilt \\
Q7: Moving/speaking slowly or fidgety & (5) Psychomotor change \\
Q8: Trouble concentrating & (8) Diminished ability to think \\
Q9: Thoughts of self-harm & (9) Suicidal ideation \\
\bottomrule
\end{tabular}
\end{table}

We test four primary hypotheses, each reflecting a link between acoustic patterns and DSM-5 indicators, particularly (5) psychomotor change and (8) diminished ability to think or concentrate. These two indicators were selected because they are consistently observable in speech and can be approximated with passive sensing in household settings. The tested directions are:

\begin{itemize}
    \item \textbf{H1 – Pitch variability}: Reduced variability of the fundamental frequency ($F_0$ standard deviation or range) is expected in individuals with psychomotor retardation or cognitive slowing, as their speech becomes monotonous and less expressive. We hypothesize a negative association between $F_0$ variability and indicators (5) and (8). 
    
    \item \textbf{H2 – Pausing behavior}: Longer or more frequent pauses reflect slowed thought processes and difficulties in speech initiation. We hypothesize a positive association between pause duration/frequency and indicators (5) and (8). 
    
    \item \textbf{H3 – Energy dynamics}: Flattened or unstable energy trajectories indicate lower vocal effort and reduced prosodic emphasis, both characteristic of psychomotor change and concentration problems. We hypothesize a negative association between energy variability and indicators (5)/(8). 
    
    \item \textbf{H4 – Speech tempo}: Slower articulation rate or reduced overall speech rate captures psychomotor retardation and reduced fluency. We hypothesize a negative association between speech rate/articulation and indicators (5)/(8). 
\end{itemize}

\subsection{Feature/Indicator Associations}
\label{sec:evaluation:results}

\begin{figure}[h]
    \centering
    \includegraphics[width=\linewidth]{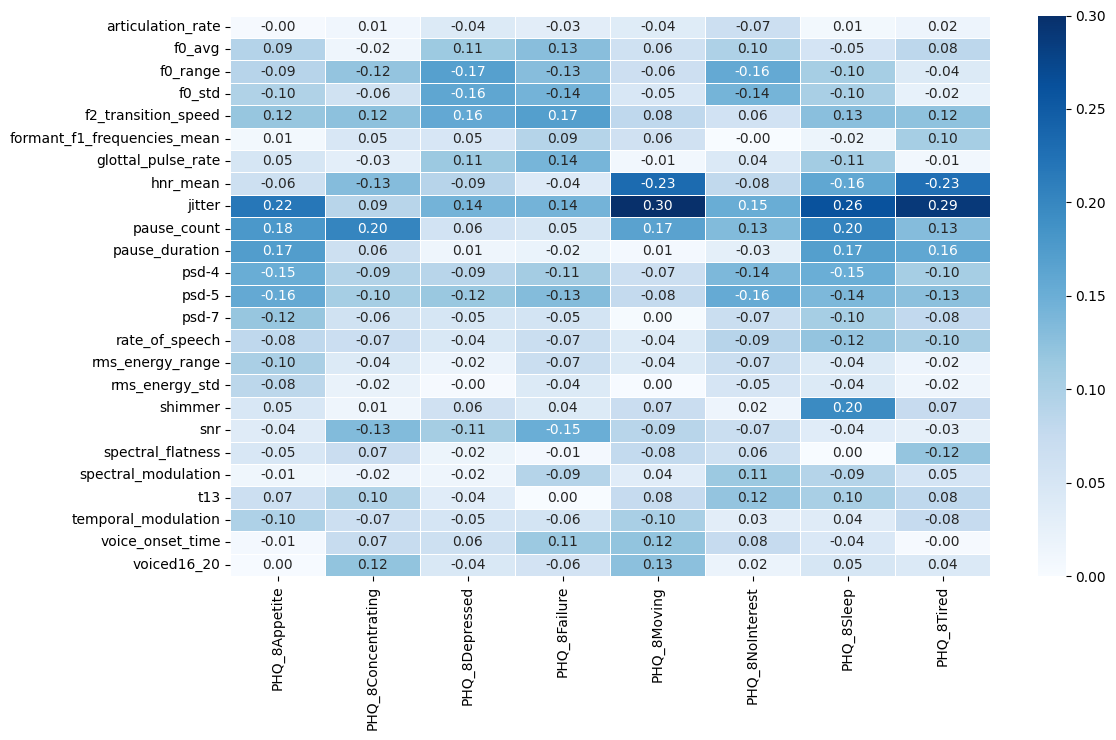}
    \caption{Effect sizes ($r$) for feature-indicator pairs on DAIC-WOZ (darker is stronger).}
    \label{fig:heatmap_r}
\end{figure}
\begin{figure}[t]
    \centering
    \includegraphics[width=\linewidth]{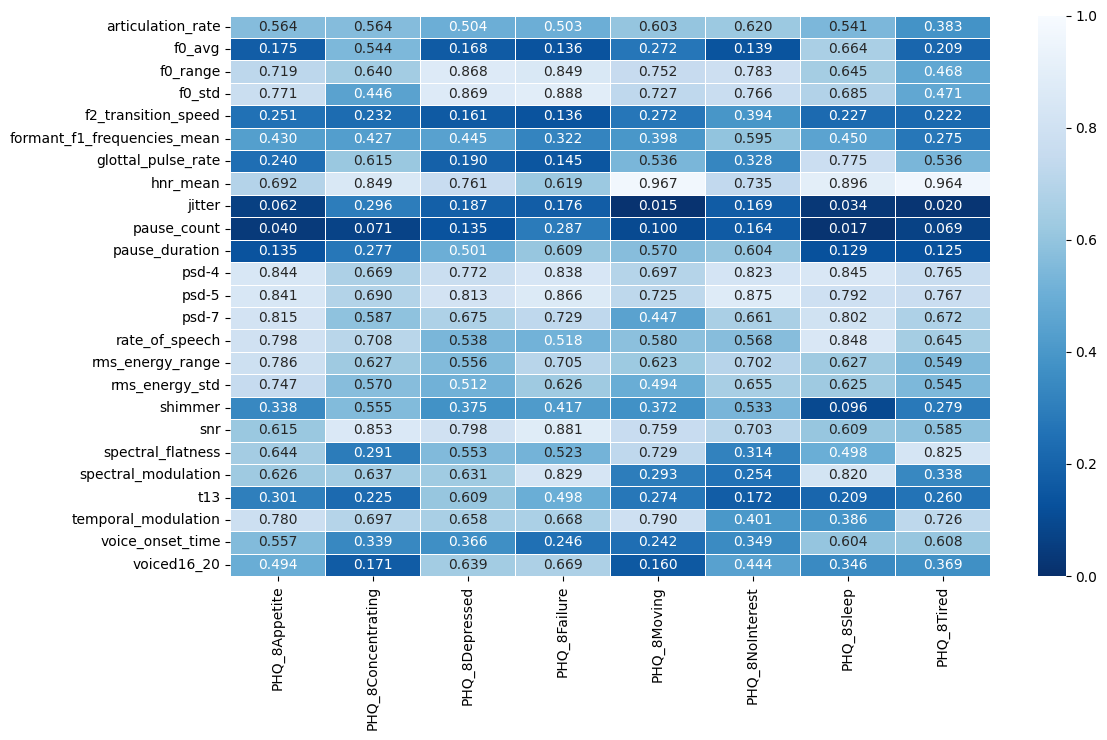}
    \caption{Significance after FDR (white=non‑sig).}
    \label{fig:heatmap_p}
\end{figure}

Figure~\ref{fig:heatmap_r} summarizes effect sizes ($r$) for each tested feature–indicator pair, and Figure~\ref{fig:heatmap_p} shows the corresponding p-values after False Discovery Rate (FDR) control. Darker cells in Figure~\ref{fig:heatmap_r} indicate stronger correlations, while white cells in Figure~\ref{fig:heatmap_p} highlight non-significant results after correction.

\begin{table}[t]
\centering
\caption{Hypothesis, directionality and robustness.}
\label{tab:hyp_outcomes}
\resizebox{\linewidth}{!}{%
\begin{tabular}{p{0.6cm}p{2.6cm}p{1.6cm}p{2.1cm}p{2.0cm}}
\toprule
\textbf{H} & \textbf{Feature group} & \textbf{DSM-5 Ind.} & \textbf{Dir.\ (LF)} & \textbf{Outcome} \\
\midrule
H1 & $F_0$ var./range & (5),(8) & negative & supported \\
H2 & Pause dur./freq. & (5),(8) & positive & supported \\
H3 & Energy dynamics & (5),(8) & negative & partial support \\
H4 & Speech rate/artic. & (5),(8) & negative & supported \\
\bottomrule
\end{tabular}%
}
\end{table}

Overall, the observed patterns align with the four hypotheses (\cf Table~\ref{tab:hyp_outcomes}). None remain statistically significant after FDR correction, reflecting the limited sample size of DAIC-WOZ and the multifactorial nature of depression. Reduced $F_0$ variability (H1) correlates negatively with psychomotor change and concentration difficulty; longer pauses (H2) correlate positively with the same indicators; energy dynamics (H3) show weaker, inconsistent negative trends; and slower speech tempo (H4) correlates negatively with both indicators, consistent with psychomotor retardation.

\subsection{Explainability}
\label{sec:evaluation:explainability}

A key property of \textit{IHearYou} is that it makes explicit how acoustic features contribute to DSM-5 indicators. The LF encodes these relations, so each processing stage—raw features, aggregation, contextualization, and indicator scoring—can be inspected. This enables users to see how evidence accumulates and which indicators share support from the same feature. 

\begin{figure}[h]
    \centering
    \includegraphics[width=0.9\linewidth]{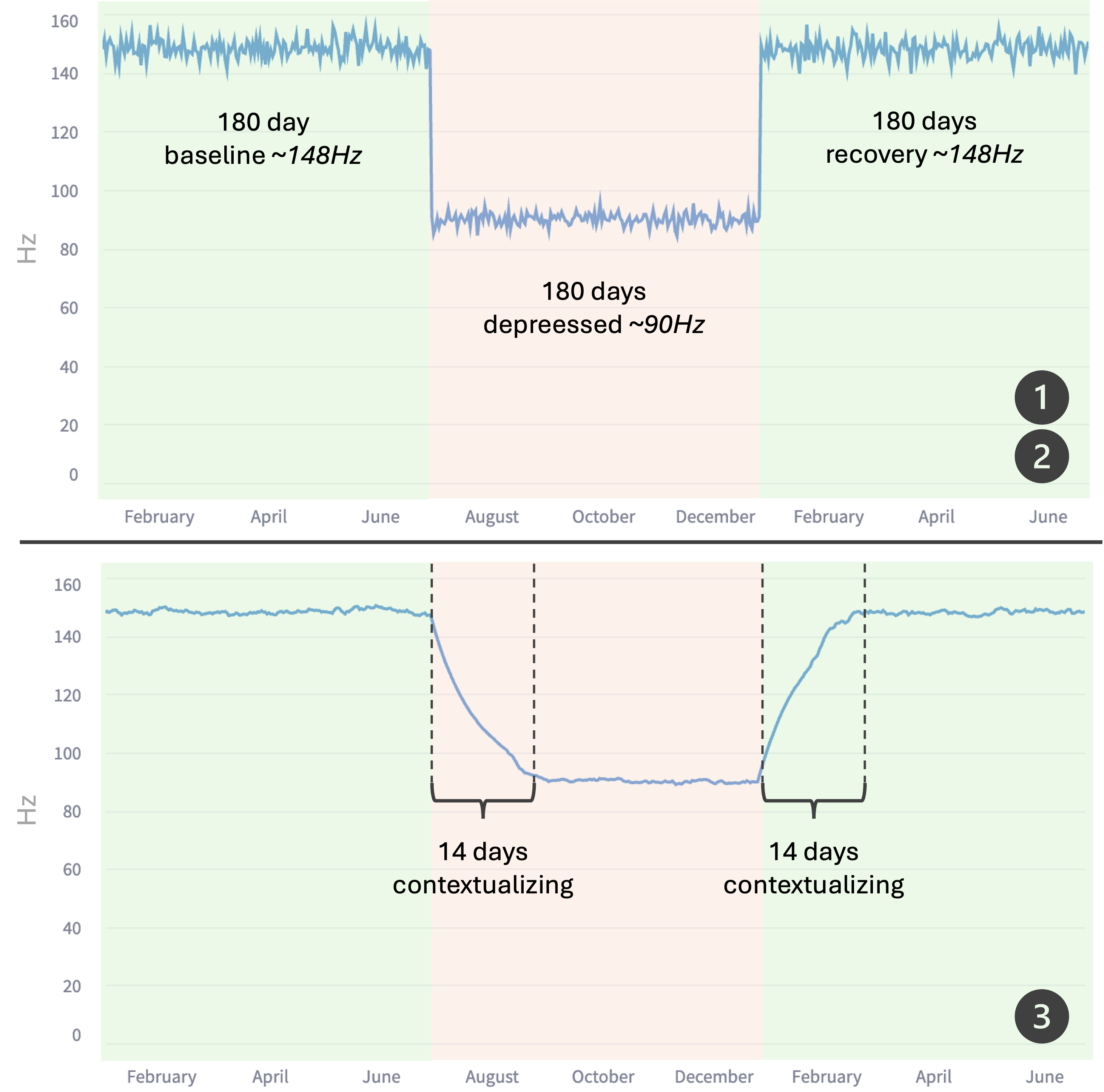}
    \includegraphics[width=0.9\linewidth]{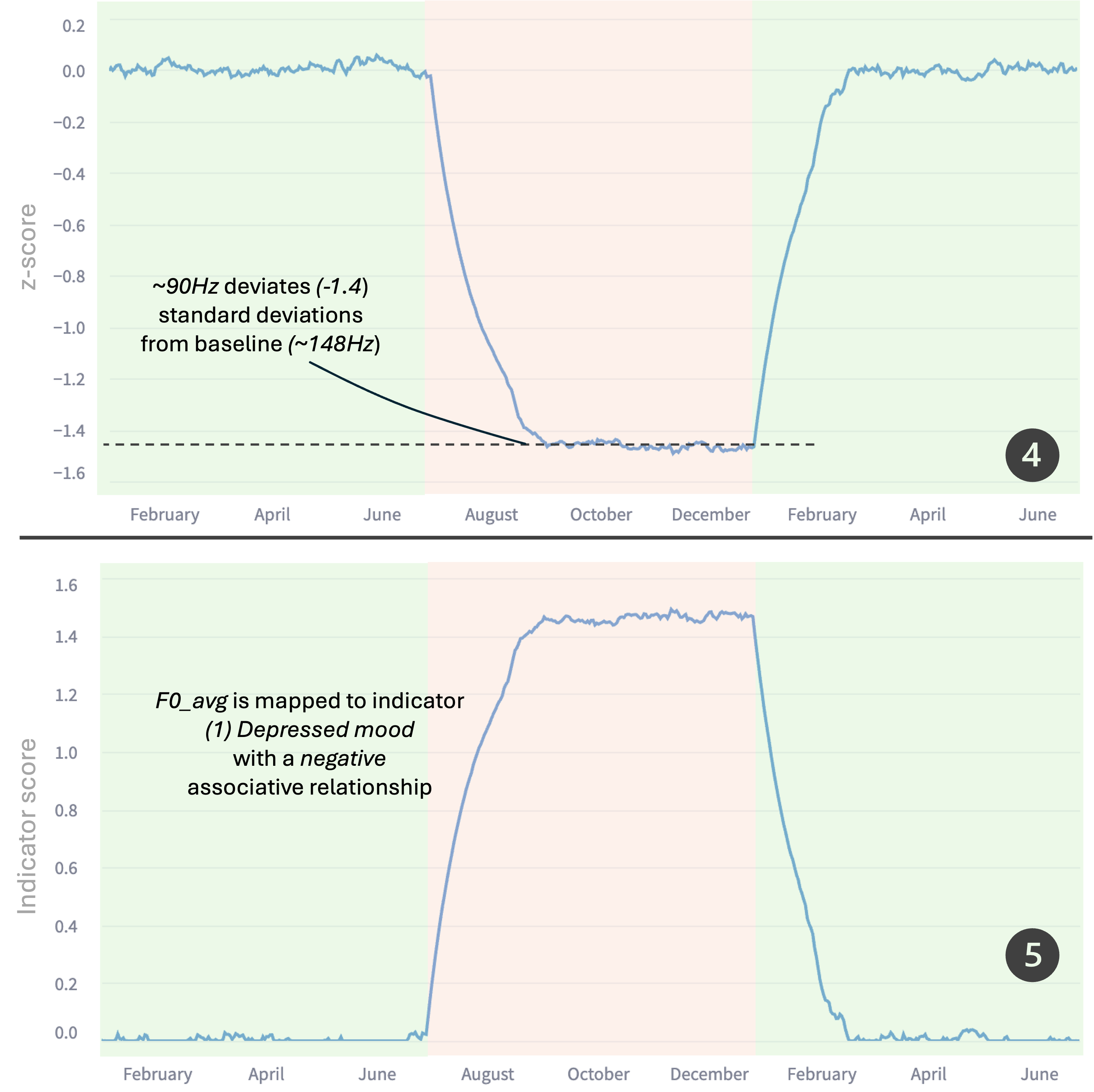}
     \caption{Pass-through with simulated $F_0$ variability. Each panel shows how changes propagate through the pipeline stages, resulting in DSM-5 indicator scores across baseline, depressed, and recovery phases.}
    \label{fig:simulated_pass_through}
\end{figure}

Figure~\ref{fig:simulated_pass_through} illustrates this step-wise traceability. Instead of returning only a global “depressed/not depressed” label, the system shows how changes in $F_0$ variability affect a specific indicator (psychomotor change). This provides interpretable rationales that clinicians or researchers can review and contest.


\subsection{Temporal Behavior (Streaming Audio)}
\label{sec:evaluation:temporal}

\begin{figure}[t]
    \centering
    \includegraphics[width=\linewidth]{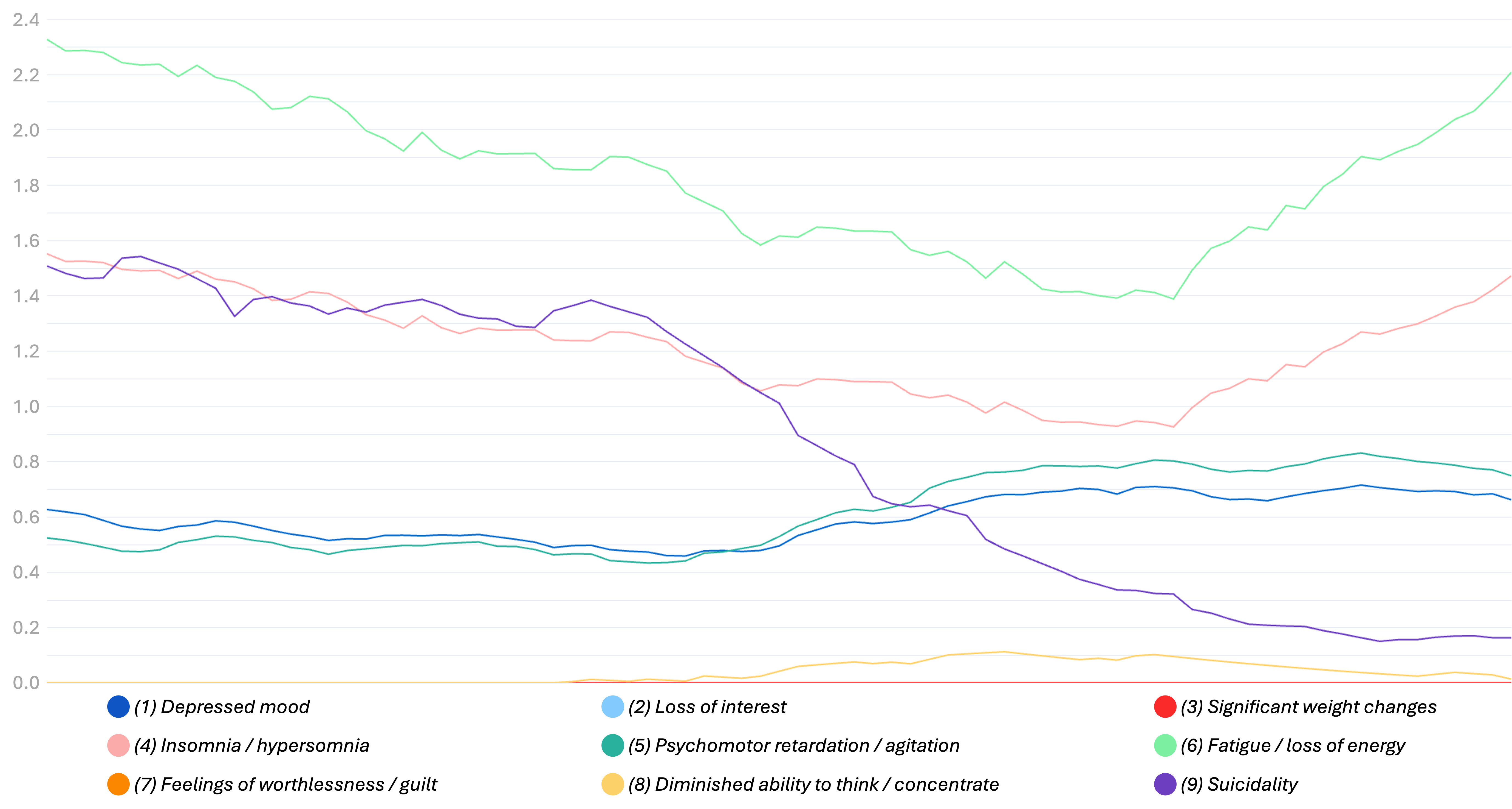}
    \caption{Indicator score trajectories during simulated streaming (10\,s windows, EMA as configured).}
    \label{fig:stream_timeline}
\end{figure}

To illustrate how the LF behaves under streaming conditions, Figure~\ref{fig:stream_timeline} plots DSM-5 indicator trajectories for a controlled input sequence. We used the Toronto Emotional Speech Set (TESS) \cite{TESS_dataset}, concatenating all available speech samples for one individual participant, such that two \textit{wav} files are created: the first contains only \textit{non-depressed} speech (\textit{long\_nondepressed\_sample\_nobreak.wav}), while the second contains \textit{depressed} speech (\textit{long\_depressed\_sample\_nobreak.wav}) \cite{anon2025blinded}. The synthetic input contained segments with slower articulation, reduced prosody, and inserted pauses, allowing us to observe whether the indicator scores evolve consistently with our hypotheses.

Figure~\ref{fig:stream_timeline} show that sustained slow speech and reduced prosodic variability gradually increase the score for indicator (5) psychomotor change, consistent with \textbf{H4} (speech tempo) and \textbf{H1} (pitch variability). When a transient long pause occurs, the EMA filter damps the spike rather than triggering a full indicator activation, matching the DSM-5 requirement for symptom persistence and supporting \textbf{H2} (pausing behavior). Energy fluctuations are also smoothed over time, indicating that short bursts of loudness do not dominate the output, which is consistent with the partial support for \textbf{H3} (energy dynamics) observed in the corpus study.

\subsection{Efficiency on Edge Hardware}
\label{sec:evaluation:efficiency}

\begin{figure}[b]
    \centering
    \includegraphics[width=1.0\linewidth]{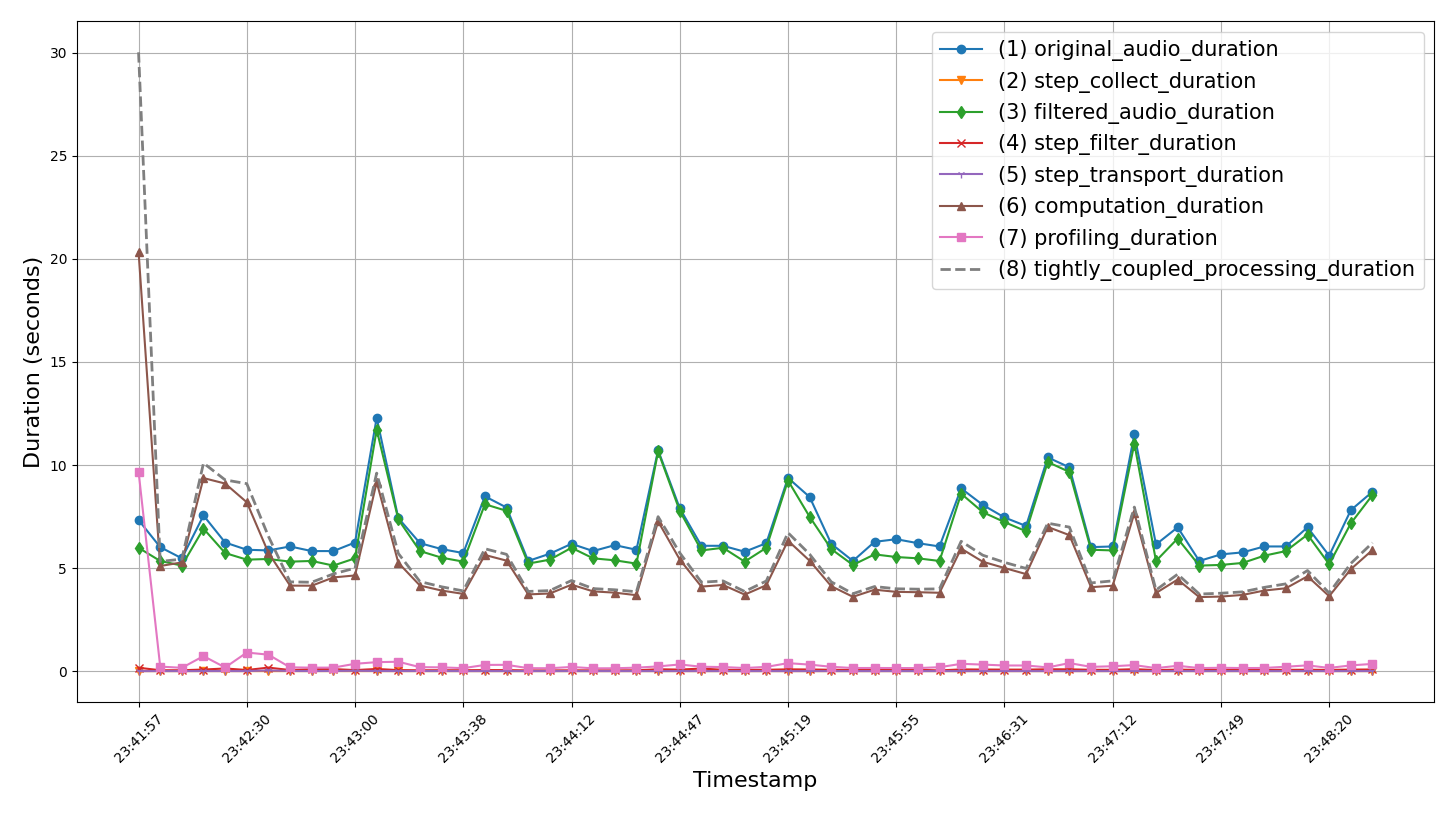}
    \caption{Processing duration measurements under continuous audio load. }
    \label{fig:performance_test_plot}
\end{figure}

A central goal of \textit{IHearYou} is to run entirely on household edge devices, without transmitting raw voice data to the cloud. We benchmarked the pipeline on an edge-class device (MacBook Pro M1 with 16 GB RAM). using continuous audio from the TESS dataset. Processing times were measured at each stage of the data pipeline and compared against the real-time constraint. Figure~\ref{fig:performance_test_plot} shows that although the first segments required longer than their audio duration, the system quickly stabilized. 

\begin{figure}[t]
    \centering
    \includegraphics[width=1.0\linewidth]{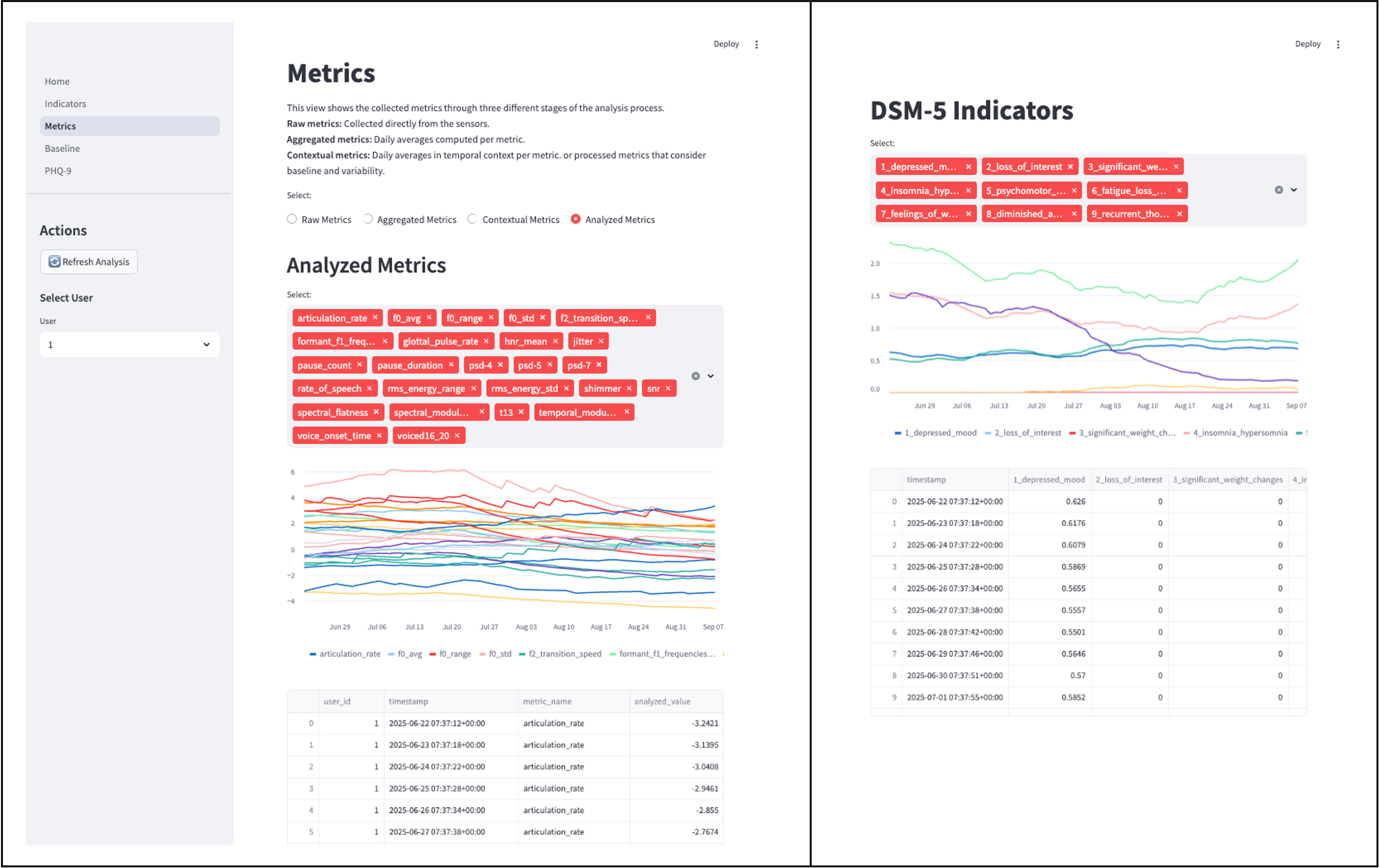}
    \caption{Web dashboard interface illustrating (left) raw metrics captured from audio input and (right) indicator derivation.}
    \label{fig:dashboard_screenshots_1}
\end{figure}

The system is implemented end-to-end, including a web-based dashboard that exposes each stage of the analysis in an interpretable form. The dashboard enables users to inspect raw metrics, contextualized features, and derived DSM-5 indicator scores, and to submit PHQ-9 responses for baseline calibration. Figures~\ref{fig:dashboard_screenshots_1} illustrate selected views of the dashboard, showing that the \textit{IHearYou} system is not only a conceptual framework but a practical tool that can operate locally. In addition, the source-code is available \cite{depressiondetection2025GitHubRepo} to foster reproducibility and further research.

\subsection{Discussion and Limitations}
\label{section:evaluation:discussion}


\textbf{Indicator-level interpretability.} The LF defines explicit feature–indicator mappings and enforces temporal and decision rules consistent with DSM-5. Because the mapping is many-to-many, a single acoustic feature can contribute to multiple indicators (\eg reduced $F_0$ variability affects both depressed mood and concentration). Rather than reducing this to a single global score, the system makes such overlaps explicit, allowing to see which indicators share evidence and why. The evaluation shows directionally consistent associations for \textbf{H1} (pitch variability), \textbf{H2} (pausing behavior), and \textbf{H4} (speech tempo), with partial support for \textbf{H3} (energy dynamics), across DAIC-WOZ heatmaps (Figures~\ref{fig:heatmap_r}--\ref{fig:heatmap_p}) and the streaming simulation (Figure~\ref{fig:stream_timeline}). 

\textbf{Edge-first, privacy-preserving operation.} The design runs locally so that raw speech remains local. This is an essential characteristic for sensitive family contexts and aligns with our architectural constraints. The timing study (Figure~\ref{fig:performance_test_plot}) confirms that the pipeline sustains real-time processing on edge-class hardware. The dashboard (Figure~\ref{fig:dashboard_screenshots_1}), although requiring more specific knowledge to inspect raw metrics, helps to understand contextualized features,  indicator scores without cloud offloading.

\textbf{Health-oriented small language models.} 
A next step is to add low-cost contextual signals (\eg coarse activity or sleep summaries) and fuse them late at the decision level, leaving the audio feature pipeline and privacy model unchanged. This aims at achieving a balanced trade-off between accuracy and compute/memory while keeping the system explainable and auditable. Compact domain-specific small language models (\eg MedGemma \cite{sellergren2025medgemma}) suggest a path to edge-feasible reasoning, but remains largely as a black-box model that hinders their interpretability (by doctors) and potential use as input for diagnosis.



\textbf{Practical implications.} Within these limits, an edge-deployable, indicator-level audio system is a feasible early-warning tool for households. It does not replace clinical diagnosis, but it can surface sustained changes in monotony, pausing, and speech tempo that warrant attention—especially for children and teenagers where early signs are often missed.

\section{Summary and Future Work}
\label{sec:summary}



This paper introduced \textit{IHearYou} \cite{depressiondetection2025GitHubRepo}, a framework that links acoustic features to DSM-5 indicators in an explainable manner and able to run in commodity hardware. Based on the DAIC-WOZ and TESS datasets, it was introduced directional associations between typical speech features and indicators, such as pitch variability, pausing, and speech tempo under a reproducible, FDR-controlled protocol. Future work will strengthen feature–indicator mappings with larger and longitudinal datasets, extend coverage to symptoms less evident in voice (\eg appetite change, suicidality), and explore multimodal integration of low-cost contextual signals while preserving interpretability and edge efficiency.

\bibliographystyle{IEEEtran}
\balance
\bibliography{references.bib}

@misc{who2023depression,
  author       = {World Health Organization},
  title        = {Depressive disorder (depression)},
  year         = {2023},
  month        = {March},
  url          = {https://www.who.int/news-room/fact-sheets/detail/depression},
  note         = {Accessed: 29-04-2025},
  publisher    = {World Health Organization}
}

@article{Remes2021DeterminantsDepression,
  author    = {{O. Remes, J. F. Mendes, and P. Templeton}},
  title     = {Biological, psychological, and social determinants of depression: A review of recent literature},
  journal   = {Brain Sciences},
  year      = {2021},
  volume    = {11},
  number    = {12},
  pages     = {1633},
  doi       = {10.3390/brainsci11121633},
  publisher = {MDPI},
  url       = {https://www.mdpi.com/2076-3425/11/12/1633},
  note      = {PMID: 34942936, PMCID: PMC8699555}
}

@book{ahrq2022child,
  title        = {2022 National Healthcare Quality and Disparities Report},
  author       = {{Agency for Healthcare Research and Quality (US)}},
  year         = {2022},
  month        = {October},
  address      = {Rockville (MD)},
  publisher    = {Agency for Healthcare Research and Quality (US)},
  note         = {Internet},
  url          = {https://www.ncbi.nlm.nih.gov/books/NBK587174/}
}

@article{linthicum2023ubiquitous,
  title = {Megatrend alert: The rise of ubiquitous computing},
  author = {D. Linthicum},
  journal = {InfoWorld},
  year = {2023},
  month = {July},
  day = {21},
  url = {https://www.infoworld.com/article/2338818/megatrend-alert-the-rise-of-ubiquitous-computing.html}
}

@misc{dilmegani2025tinyml,
  author       = {C. Dilmegani},
  title        = {TinyML (EdgeAI) in 2025: Machine learning at the edge},
  year         = {2025},
  url          = {https://research.aimultiple.com/tinyml/},
  note         = {Accessed: 23-05-2025}
}

@article{di2024voicepitch,
  author       = {{Y. Di, E. Rahmani, J. Mefford, et al.}},
  title        = {Unraveling the associations between voice pitch and major depressive disorder: a multisite genetic study},
  journal      = {Molecular Psychiatry},
  year         = {2024},
  doi          = {10.1038/s41380-024-02877-y},
  url          = {https://doi.org/10.1038/s41380-024-02877-y},
  note         = {Published: 31 December 2024},
}

@INPROCEEDINGS{cohn2009detecting,
  author={{J. F. Cohn, T. S. Kruez, I. Matthews, Y. Yang, M. H. Nguyen, M. T. Padilla, F. Zhou, and F. De la Torre}},
  booktitle={2009 3rd International Conference on Affective Computing and Intelligent Interaction and Workshops}, 
  title={Detecting depression from facial actions and vocal prosody}, 
  year={2009},
  volume={},
  number={},
  pages={1-7},
  doi={10.1109/ACII.2009.5349358}
}

@article{cummins2015review,
    title = {A review of depression and suicide risk assessment using speech analysis},
    journal = {Speech Communication},
    volume = {71},
    pages = {10-49},
    year = {2015},
    issn = {0167-6393},
    doi = {https://doi.org/10.1016/j.specom.2015.03.004},
    url = {https://www.sciencedirect.com/science/article/pii/S0167639315000369},
    author = {{N. Cummins, S. Scherer, J. Krajewski, S. Schnieder, J. Epps, and T. F. Quatieri}}
}

@article{baumeister2007behavior,
    author = {{R. F. Baumeister, K. D. Vohs, and D. C. Funder}},
    title ={Psychology as the science of self-reports and finger movements: Whatever happened to actual behavior?},
    journal = {Perspectives on Psychological Science},
    volume = {2},
    number = {4},
    pages = {396-403},
    year = {2007},
    doi = {10.1111/j.1745-6916.2007.00051.x},
    note ={PMID: 26151975},
    URL = {https://doi.org/10.1111/j.1745-6916.2007.00051.x},
    eprint = {https://doi.org/10.1111/j.1745-6916.2007.00051.x},
}

@article{AdamsQuackenbush2019,
    author = {{N. M. Adams-Quackenbush, R. Horselenberg, J. Hubert, A. Vrij, and P. van Koppen}},
    title = {Interview expectancies: awareness of potential biases influences behaviour in interviewees},
    journal = {Psychiatry, Psychology and Law},
    volume = {26},
    number = {1},
    pages = {150--166},
    year = {2019},
    publisher = {Routledge},
    doi = {10.1080/13218719.2018.1485522},
    note ={PMID: 31984070},
    URL = {https://doi.org/10.1080/13218719.2018.1485522},
    eprint = {https://doi.org/10.1080/13218719.2018.1485522}
}

@article{LIU202544,
    title = {Multimodal depression recognition and analysis: Facial expression and body posture changes via emotional stimuli},
    journal = {Journal of Affective Disorders},
    volume = {381},
    pages = {44-54},
    year = {2025},
    issn = {0165-0327},
    doi = {https://doi.org/10.1016/j.jad.2025.03.155},
    url = {https://www.sciencedirect.com/science/article/pii/S0165032725005038},
    author = {{Y. Liu, X. Li, M. Wang, J. Bi, S. Lin, Q. Wang, Y. Yu, J. Ye, and Y. Zheng}},
}

@article{herm2023mlexplainability,
    title = {Stop ordering machine learning algorithms by their explainability! A user-centered investigation of performance and explainability},
    journal = {International Journal of Information Management},
    volume = {69},
    pages = {102538},
    year = {2023},
    issn = {0268-4012},
    doi = {https://doi.org/10.1016/j.ijinfomgt.2022.102538},
    url = {https://www.sciencedirect.com/science/article/pii/S026840122200072X},
    author = {{L.-V. Herm, K. Heinrich, J. Wanner, and C. Janiesch}},
}

@article{kraepelin1921manic,
  author       = {E. Kraepelin},
  title        = {Manic depressive insanity and paranoia},
  journal      = {The Journal of Nervous and Mental Disease},
  volume       = {53},
  number       = {4},
  pages        = {350},
  year         = {1921},
  month        = apr,
  publisher    = {Lippincott Williams \& Wilkins}
}

@article{fried2025depressionsumscores,
  author    = {{I. F. Eiko and R. M. Nesse}},
  title     = {Depression sum-scores don’t add up: why analyzing specific depression symptoms is essential},
  journal   = {BMC Medicine},
  volume    = {13},
  number    = {1},
  pages     = {72},
  year      = {2015},
  doi       = {10.1186/s12916-015-0325-4},
  url       = {https://doi.org/10.1186/s12916-015-0325-4},
  publisher = {BioMed Central},
  published = {2015-04-06}
}

@article{Min2023detectingDepressionVideoLogs,
  author    = {{K. Min, J. Yoon, M. Kang, D. Lee, E. Park, and J. Han}},
  title     = {Detecting depression on video logs using audiovisual features},
  journal   = {Humanities and Social Sciences Communications},
  volume    = {10},
  number    = {1},
  year      = {2023},
  pages     = {788},
  doi       = {10.1057/s41599-023-02313-6},
  url       = {https://doi.org/10.1057/s41599-023-02313-6}
}

@incollection{jahan2022multimodality,
    title = {Chapter two - multimodal depression detection using machine learning},
    editor = {{S. Jain, K. Pandeym, P. Jain, and K. P. Seng}},
    booktitle = {Artificial intelligence, machine learning, and mental health in pandemics},
    publisher = {Academic Press},
    pages = {53-72},
    year = {2022},
    isbn = {978-0-323-91196-2},
    doi = {https://doi.org/10.1016/B978-0-323-91196-2.00005-3},
    url = {https://www.sciencedirect.com/science/article/pii/B9780323911962000053},
    author = {{R. Jahan and M. M. Tripathi}},
}

@article{sadeghi2024harnessing,
  author    = {{M. Sadeghi, R. Richer, B. Egger, et al.}},
  title     = {Harnessing multimodal approaches for depression detection using large language models and facial expressions},
  journal   = {npj Mental Health Research},
  volume    = {3},
  pages     = {66},
  year      = {2024},
  doi       = {10.1038/s44184-024-00112-8},
  url       = {https://doi.org/10.1038/s44184-024-00112-8},
  publisher = {Nature Portfolio}
}

@misc{norvig_spell_corrector,
  author       = {P. Norvig},
  title        = {How to write a spelling corrector},
  howpublished = {Online},
  year         = {2007},
  note         = {Accessed: 26-06-2025},
  url          = {http://norvig.com/spell-correct.html},
}

@article{Chiong2021smn,
    title = {A textual-based featuring approach for depression detection using machine learning classifiers and social media texts},
    journal = {Computers in Biology and Medicine},
    volume = {135},
    pages = {104499},
    year = {2021},
    issn = {0010-4825},
    doi = {https://doi.org/10.1016/j.compbiomed.2021.104499},
    url = {https://www.sciencedirect.com/science/article/pii/S0010482521002936},
    author = {{R. Chiong, G. S. Budhi, S. Dhakal, and F. Chiong}},
}

@inproceedings{shen2017dd,
  author    = {{G. Shen, J. Jia, L. Nie, F. Feng, C. Zhang, T. Hu, T.-S. Chua, and W. Zhu}},
  title     = {Depression detection via harvesting social media: A multimodal dictionary learning solution},
  booktitle = {Proceedings of the twenty-sixth International Joint Conference on
               Artificial Intelligence, {IJCAI-17}},
  pages     = {3838--3844},
  year      = {2017},
  doi       = {10.24963/ijcai.2017/536},
  url       = {https://doi.org/10.24963/ijcai.2017/536},
}

@article{Cacheda2019earlyDD,
  author    = {{F. Cacheda, D. Fernandez, F. J. Novoa, and V. Carneiro}},
  title     = {Early detection of depression: Social network analysis and random forest techniques},
  journal   = {Journal of Medical Internet Research},
  volume    = {21},
  number    = {6},
  pages     = {e12554},
  year      = {2019},
  doi       = {10.2196/12554},
  url       = {https://doi.org/10.2196/12554}
}

@article{Kong2022,
  author  = {{X. Kong, Y. Yao, C. Wang, Y. Wang, J. Teng, and X. Qi}},
  title   = {Automatic identification of depression using facial images with deep convolutional neural network},
  journal = {Medical Science Monitor, International Medical Journal of Experimental and Clinical Research},
  year    = {2022},
  volume  = {28},
  pages   = {e936409},
  doi     = {10.12659/MSM.936409},
  url     = {https://doi.org/10.12659/MSM.936409}
}

@article{tufail2023dd,
    title = {Depression detection with convolutional neural networks: A step towards improved mental health care},
    journal = {Procedia Computer Science},
    volume = {224},
    pages = {544-549},
    year = {2023},
    issn = {1877-0509},
    doi = {https://doi.org/10.1016/j.procs.2023.09.079},
    url = {https://www.sciencedirect.com/science/article/pii/S1877050923011262},
    author = {{H. Tufail, S. M. Cheema, M. Ali, I. M. Pires, and N. M. Garcia}},
}

@article{sharma2020twitter,
    author = {{A. Sharma and U. Ghose}},
    year = {2020},
    month = {01},
    pages = {325-334},
    title = {Sentimental analysis of twitter data with respect to general elections in India},
    volume = {173},
    journal = {Procedia Computer Science},
    doi = {10.1016/j.procs.2020.06.038}
}

@book{Bains2023,
  title        = {Major depressive disorder},
  author       = {{N. Bains and S. Abdijadid}},
  year         = {2023},
  publisher    = {StatPearls Publishing},
  address      = {Treasure Island (FL)},
  note         = {Updated 2023 April 10},
  url          = {https://www.ncbi.nlm.nih.gov/books/NBK559078/}
}

@misc{depressiondetection2025GitHubRepo,
  author       = {Jonas Länzlinger},
  title        = {{Depression Detection GitHub Repository}},
  year         = {2025},
  howpublished = {\url{https://github.com/jonaslanzlinger/depression-detection}},
  note         = {Accessed: 21-06-2025}
}

@inproceedings{anon2025blinded,
  author    = {Blinded for review},
  title     = {Blinded for review},
  booktitle = {Blinded for review},
  year      = {2025}
}

@article{Donaghy2024VoiceBiomarkers,
  author    = {{P. Donaghy, E. Ennis, M. Mulvenna, R. R. Bond, N. Kennedy, M. McTear, H. O’Connell, N. Blaylock, R. Brueckner}},
  title     = {A review of studies using machine learning to detect voice biomarkers for depression},
  journal   = {Journal of Technology in Behavioral Science},
  year      = {2024},
  month     = dec,
  day       = {12},
  volume    = {—},
  number    = {—},
  pages     = {1--15},
  doi       = {10.1007/s41347-024-00454-2},
  url       = {https://doi.org/10.1007/s41347-024-00454-2},
}

@article{Low2020keyacousticfeatures,
    author = {{D. M. Low, K. H. Bentley, S. S. Ghosh}},
    title = {Automated assessment of psychiatric disorders using speech: A systematic review},
    journal = {Laryngoscope Investigative Otolaryngology},
    volume = {5},
    number = {1},
    pages = {96-116},
    doi = {https://doi.org/10.1002/lio2.354},
    url = {https://onlinelibrary.wiley.com/doi/abs/10.1002/lio2.354},
    eprint = {https://onlinelibrary.wiley.com/doi/pdf/10.1002/lio2.354},
    year = {2020}
}

@data{TESS_dataset,
    author = {{M. K. Pichora-Fuller and K. Dupuis}},
    publisher = {Borealis},
    title = {{Toronto emotional speech set (TESS)}},
    year = {2020},
    version = {V1},
    doi = {10.5683/SP2/E8H2MF},
    url = {https://doi.org/10.5683/SP2/E8H2MF}
}

@misc{NU_CohensD_2025,
  title        = {Cohen's d},
  author       = {{National University Library}},
  howpublished = {Online statistics resource},
  month        = jun,
  year         = {2025},
  note         = {Accessed: 16-06-2025},
  url          = {https://resources.nu.edu/statsresources/cohensd}
}

@article{VANDANA2023100587,
    title = {A hybrid model for depression detection using deep learning},
    journal = {Measurement: Sensors},
    volume = {25},
    pages = {100587},
    year = {2023},
    issn = {2665-9174},
    doi = {https://doi.org/10.1016/j.measen.2022.100587},
    url = {https://www.sciencedirect.com/science/article/pii/S2665917422002215},
    author = {{Vandana, N. Marriwala, and D. Chaudhary}}
}

@article{huang2024depression,
  author = {{X. Huang, F. Wang, Y. Gao, et al.}},
  title = {Depression recognition using voice-based pre-training model},
  journal = {Scientific Reports},
  volume = {14},
  pages = {12734},
  year = {2024},
  publisher = {Nature Publishing Group},
  doi = {10.1038/s41598-024-63556-0},
  url = {https://doi.org/10.1038/s41598-024-63556-0}
}

@INPROCEEDINGS{Govindasamy2021,
  author={{K. A. L. Govindasamy and N. Palanichamy}},
  booktitle={2021 5th International Conference on Intelligent Computing and Control Systems (ICICCS)}, 
  title={Depression detection using machine learning techniques on Twitter data}, 
  year={2021},
  volume={},
  number={},
  pages={960-966},
  doi={10.1109/ICICCS51141.2021.9432203}
}

@article{mao2023systematic,
  author = {{K. Mao, Y. Wu, and J. Chen}},
  title = {A systematic review on automated clinical depression diagnosis},
  journal = {npj Mental Health Research},
  volume = {2},
  pages = {20},
  year = {2023},
  doi = {10.1038/s44184-023-00040-z}
}

@book{bird2009nltk,
  author    = {{S. Bird, E. Klein, and E. Loper}},
  title     = {Natural language processing with Python},
  publisher = {O’Reilly Media, Inc.},
  year      = {2009},
  url       = {https://www.nltk.org/book/}
}

@misc{usc_daic_woz,
  title        = {DAIC‑WOZ database \& extended DAIC database},
  author       = {USC Institute for Creative Technologies},
  howpublished = {\url{https://dcapswoz.ict.usc.edu/}},
  year         = {2025},
  note         = {Accessed: 10-06-2025},
  month        = jun,
}

@article{sellergren2025medgemma,
  title={Medgemma technical report},
  author={Sellergren, Andrew and Kazemzadeh, Sahar and Jaroensri, Tiam and Kiraly, Atilla and Traverse, Madeleine and Kohlberger, Timo and Xu, Shawn and Jamil, Fayaz and Hughes, C{\'\i}an and Lau, Charles and others},
  journal={arXiv preprint arXiv:2507.05201},
  year={2025}
}

@manual{americanpsychiatricassociation2022dsm5,
    author = {American Psychiatric Association},
    title = {Diagnostic and statistical manual of mental disorders: DSM-5-TR},
    edition = {Fifth Edition},
    address = {Washington, DC},
    publisher = {American Psychiatric Association Publishing},
    year = {2022}
}

@inproceedings{lin2020sensemood,
    author = {{C. Lin, P. Hu, H. Su, S. Li, J. Mei, J. Zhou, H. Leung}},
    title = {SenseMood: Depression detection on social media},
    year = {2020},
    isbn = {9781450370875},
    publisher = {Association for Computing Machinery},
    address = {New York, NY, USA},
    url = {https://doi.org/10.1145/3372278.3391932},
    doi = {10.1145/3372278.3391932},
    booktitle = {Proceedings of the 2020 International Conference on Multimedia Retrieval},
    pages = {407–411},
    numpages = {5},
    location = {Dublin, Ireland},
    series = {ICMR '20}
}

@article{Yadav2022ADD,
    author = {{U. Yadav, A. Sharma, and D. Patil}},
    year = {2022},
    month = {10},
    pages = {},
    title = {Review of automated depression detection: Social posts, audio and video, open challenges and future direction},
    volume = {35},
    journal = {Concurrency and Computation: Practice and Experience},
    doi = {10.1002/cpe.7407}
}

@article{kroenke2001phq9,
  title     = {The PHQ-9: validity of a brief depression severity measure},
  author    = {{K. Kroenke, R. L. Spitzer, and J. B. W. Williams}},
  journal   = {Journal of General Internal Medicine},
  volume    = {16},
  number    = {9},
  pages     = {606--613},
  year      = {2001},
  month     = {September},
  doi       = {10.1046/j.1525-1497.2001.016009606.x},
  pmid      = {11556941},
  pmcid     = {PMC1495268},
  publisher = {Springer}
}

@article{kroenke2009phq8,
  title = {The PHQ-8 as a measure of current depression in the general population},
  author = {{K. Kroenke, T. W. Strine, R. L. Spitzer, J. B. Williams, J. T. Berry, and A. H. Mokdad}},
  journal = {Journal of Affective Disorders},
  volume = {114},
  number = {1-3},
  pages = {163--173},
  year = {2009},
  doi = {10.1016/j.jad.2008.06.026}
}

@article{trisatmaja2022bimodalvoice,
    author = {{B. T. Atmaja, A. Sasou, and M. Akagi}},
    year = {2022},
    month = {03},
    pages = {},
    title = {Survey on bimodal speech emotion recognition from acoustic and linguistic information fusion},
    volume = {140},
    journal = {Speech Communication},
    doi = {10.1016/j.specom.2022.03.002}
}

@misc{SileroVAD,
  author = {Silero team},
  title = {Silero VAD: pre-trained enterprise-grade voice activity detector (VAD), number detector and language classifier},
  year = {2024},
  publisher = {GitHub},
  journal = {GitHub repository},
  howpublished = {\url{https://github.com/snakers4/silero-vad}},
}

@article{zhang2024multimodal,
  title={Multimodal sensing for depression risk detection: Integrating audio, video, and text data},
  author={Zhang, Zhenwei and Zhang, Shengming and Ni, Dong and Wei, Zhaoguo and Yang, Kongjun and Jin, Shan and Huang, Gan and Liang, Zhen and Zhang, Li and Li, Linling and others},
  journal={Sensors},
  volume={24},
  number={12},
  pages={3714},
  year={2024},
  publisher={MDPI}
}

\end{document}